\documentstyle [10pt,amsfonts] {article}
\input epsf

\newcommand{\beq}{\begin{equation}}
\newcommand{\eeq}{\end{equation}}
\newcommand{\beqs}{\begin{eqnarray}}
\newcommand{\eeqs}{\end{eqnarray}}

\catcode`@=11
\@addtoreset{equation}{section}
\@addtoreset{equation}{subsection}
\def\theequation{\ifnum\value{section}=0 \arabic{equation}\ignorespaces
\else \ifnum\value{section}=-1 A.\arabic{equation}\ignorespaces
\else \ifnum\value{subsection}=0 \thesection.\arabic{equation}\ignorespaces
\else \thesection.\arabic{subsection}.\arabic{equation}\ignorespaces
                           \fi
                      \fi
                 \fi}
\catcode`@=12

\begin{document}

\def\thefootnote{\fnsymbol{footnote}}

\baselineskip 5.0mm

\vspace{4mm}

\begin{center}

{\Large \bf Exact Potts Model Partition Functions on Strips of the Honeycomb
Lattice}

\vspace{8mm}

\setcounter{footnote}{0}
Shu-Chiuan Chang\footnote{email: shu-chiuan.chang@sunysb.edu} and 
\setcounter{footnote}{6}
Robert Shrock\footnote{email: robert.shrock@sunysb.edu}

\vspace{6mm}

C. N. Yang Institute for Theoretical Physics  \\
State University of New York       \\
Stony Brook, N. Y. 11794-3840  \\

\vspace{10mm}

{\bf Abstract}
\end{center}

We present exact calculations of the partition function of the $q$-state Potts
model on (i) open, (ii) cyclic, and (iii) M\"obius strips of the honeycomb
(brick) lattice of width $L_y=2$ and arbitrarily great length.  In the
infinite-length limit the thermodynamic properties are discussed.  The
continuous locus of singularities of the free energy is determined in the $q$
plane for fixed temperature and in the complex temperature plane for fixed $q$
values.  We also give exact calculations of the zero-temperature partition
function (chromatic polynomial) and $W(q)$, the exponent of the ground-state
entropy, for the Potts antiferromagnet for honeycomb strips of type (iv)
$L_y=3$, cyclic, (v) $L_y=3$, M\"obius, (vi) $L_y=4$, cylindrical, and (vii)
$L_y=4$, open.  In the infinite-length limit we calculate $W(q)$ and determine
the continuous locus of points where it is nonanalytic.  We show that our exact
calculation of the entropy for the $L_y=4$ strip with cylindrical boundary
conditions provides an extremely accurate approximation, to a few parts in
$10^5$ for moderate $q$ values, to the entropy for the full 2D honeycomb
lattice (where the latter is determined by Monte Carlo measurements since no
exact analytic form is known).

\vspace{16mm}

\pagestyle{empty}
\newpage

\pagestyle{plain}
\pagenumbering{arabic}
\renewcommand{\thefootnote}{\arabic{footnote}}
\setcounter{footnote}{0}

\section{Introduction}

The $q$-state Potts model has served as a valuable model for the study of phase
transitions and critical phenomena \cite{potts,wurev}.  On a lattice, or, more
generally, on a (connected) graph $G$, at temperature $T$, this model is 
defined by the partition function
\beq
Z(G,q,v) = \sum_{ \{ \sigma_n \} } e^{-\beta {\cal H}}
\label{zfun}
\eeq
with the (zero-field) Hamiltonian
\beq
{\cal H} = -J \sum_{\langle i j \rangle} \delta_{\sigma_i \sigma_j}
\label{ham}
\eeq
where $\sigma_i=1,...,q$ are the spin variables on each vertex (site) 
$i \in G$;
$\beta = (k_BT)^{-1}$; and $\langle i j \rangle$ denotes pairs of adjacent
vertices.  The graph $G=G(V,E)$ is defined by its vertex set $V$ and its edge
set $E$; we denote the number of vertices of $G$ as $n=n(G)=|V|$ and the
number of edges of $G$ as $e(G)=|E|$.  We use the notation
\beq
K = \beta J \ , \quad a = u^{-1} = e^K \ , \quad v = a-1
\label{kdef}
\eeq
so that the physical ranges are (i) $a \ge 1$, i.e., $v \ge 0$ corresponding to
$\infty \ge T \ge 0$ for the Potts ferromagnet, and (ii) $0 \le a \le 1$,
i.e., $-1 \le v \le 0$, corresponding to $0 \le T \le \infty$ for the Potts
antiferromagnet.  One defines the (reduced) free energy per site $f=-\beta F$,
where $F$ is the actual free energy, via
\beq
f(\{G\},q,v) = \lim_{n \to \infty} \ln [ Z(G,q,v)^{1/n}]  \ .
\label{ef}
\eeq
where we use the symbol $\{G\}$ to denote $\lim_{n \to \infty}G$ for a given
family of graphs.  In the present context, this $n \to \infty$ limit
corresponds to the limit of infinite length for a strip graph of the honeycomb
lattice of fixed width and some prescribed boundary conditions. 

Let $G^\prime=(V,E^\prime)$ be a spanning subgraph of $G$, i.e. a subgraph
having the same vertex set $V$ and an edge set $E^\prime \subseteq E$. Then
$Z(G,q,v)$ can be written as the sum \cite{birk}-\cite{kf}
\beqs
Z(G,q,v) & = & \sum_{G^\prime \subseteq G} q^{k(G^\prime)}v^{e(G^\prime)}
\label{cluster} \cr\cr\cr
& = & \sum_{r=k(G)}^{n(G)}\sum_{s=0}^{e(G)}z_{rs} q^r v^s
\label{zpol}
\eeqs
where $k(G^\prime)$ denotes the number of connected components of $G^\prime$
and $z_{rs} \ge 0$.  Since we only consider connected graphs $G$, we have
$k(G)=1$. The formula (\ref{cluster}) enables one to generalize $q$
from ${\mathbb Z}_+$ to ${\mathbb R}_+$ (keeping $v$ in its physical range).
This generalization is sometimes denoted the random cluster model \cite{kf,fk};
here we shall use the term ``Potts model'' to include both positive integral
$q$ as in the original formulation in eqs. (\ref{zfun}) and (\ref{ham}), and
the generalization to real (or complex) $q$, via eq. (\ref{cluster}).  The
formula (\ref{cluster}) shows that $Z(G,q,v)$ is a polynomial in $q$ and $v$
(equivalently, $a$) with maximum and minimum degrees indicated in
eq. (\ref{zpol}).  The Potts model partition function on
a graph $G$ is essentially equivalent to the Tutte polynomial
\cite{tutte1}-\cite{tutte3} and Whitney rank polynomial \cite{whit},
\cite{wurev}, \cite{bbook}-\cite{boll} for this graph, as discussed in the
appendix.

\vspace{6mm}

In this paper we shall present exact calculations of the Potts model partition
function for strips of the honeycomb (equivalently, brick) lattice of width
$L_y=2$ vertices and arbitrary length equal to $m$ bricks, with boundary
conditions of the following types: (i) $(FBC_y,FBC_x)=$ free or open, (ii)
$(FBC_y,PBC_x)=$ cyclic, and (iii) $(FBC_y,TPBC_x)=$ M\"obius, where $FBC$ and
$(T)PBC$ refer to free and (twisted) periodic boundary conditions,
respectively, and the longitudinal and transverse directions are taken to be
$\hat x$ (horizontal) and $\hat y$ (vertical).  This is a natural extension to
the honeycomb lattice of our previous exact calculations of Potts model
partition functions on finite-width, arbitrary-length strips of other lattices
\cite{bcc}-\cite{ka}.  Note that while the Potts model partition functions, or
equivalently, Tutte polynomials for the limit of infinite 2D triangular and
honeycomb lattices are simply related by duality, this is not true for
finite-width strips of the triangular and honeycomb lattices, owing to boundary
effects. (Dual graphs are discussed further in the appendix.)  In addition to
these results for arbitrary temperature, we present exact calculations for the
interesting case of the zero-temperature antiferromagnet for strips of the
honeycomb lattice of the following types: (iv) $L_y=3$, cyclic, (v) $L_y=3$,
M\"obius, (vi) $L_y=4$, $(PBC_y,FBC_x)=$cylindrical, and (vii) $L_y=4$, free.

There are several motivations for these exact calculations of Potts model
partition functions for strips of the honeycomb lattice. Clearly, new exact
calculations of Potts model partition functions are of value in their own
right.  From these, in the limit of infinite-length strips, one derives exact
thermodynamic functions and can make rigorous comparisons of various properties
among strips of different lattices, e.g., square, honeycomb, and triangular.
In particular, with these exact results one can study both the $T=0$ critical
point of the $q$-state Potts ferromagnet and $T=0$ properties of the Potts
antiferromagnet.  Further, we shall use our exact results to study how, for a
given value of $q$, the ground state degeneracy per site (exponent of the
ground state entropy per site) for the Potts antiferromagnet approaches its
value for the infinite honeycomb lattice.  We shall show that the exactly
determined function for the ground state degeneracy calculated for the strip
with an infinite-length strip with a rather modest width of $L_y=4$ and
cylindrical boundary conditions is already an extremely good approximation, for
a wide range of $q$ values, to the corresponding function for the full 2D
honeycomb lattice (the latter being determined by Monte Carlo measurements).
This is very useful since no exact expression for the ground state degeneracy
per site for $q > 2$ is known for the 2D honeycomb lattice.  In addition, it
was shown \cite{a} that these calculations can give insight into the
complex-temperature phase diagram of the 2D Potts model on the given lattice,
which is again useful since, except for the $q=2$ Ising case, the latter phase
diagrams are not known exactly. Finally, via the correspondence with the Tutte
polynomial, our calculations yield several quantities of relevance to
mathematical graph theory.

Various special cases of the Potts model partition function are of interest.
One special case is the zero-temperature limit of the Potts antiferromagnet. 
For sufficiently large $q$, on a given lattice or graph $G$, this 
exhibits nonzero ground state entropy (without frustration).  This is of
interest as an exception to the third law of thermodynamics \cite{al,cw}. 
This is equivalent to a ground
state degeneracy per site (vertex), $W > 1$, since $S_0 = k_B \ln W$.  The
$T=0$ (i.e., $v=-1$) partition function of the above-mentioned $q$-state Potts
antiferromagnet on $G$ satisfies
\beq 
Z(G,q,-1)=P(G,q)
\label{zp}
\eeq
where $P(G,q)$ is the chromatic polynomial (in $q$) expressing the number
of ways of coloring the vertices of the graph $G$ with $q$ colors such that no
two adjacent vertices have the same color \cite{birk,bbook,rrev,rtrev}. The 
minimum number of colors necessary for this coloring is the chromatic number
of $G$, denoted $\chi(G)$.  We have 
\beq 
W(\{G\},q)= \lim_{n \to \infty}P(G,q)^{1/n} \ .
\label{w}
\eeq
At certain special
points $q_s$ (typically $q_s=0,1,.., \chi(G)$), one has the noncommutativity of
limits
\beq
\lim_{q \to q_s} \lim_{n \to \infty} P(G,q)^{1/n} \ne \lim_{n \to
\infty} \lim_{q \to q_s}P(G,q)^{1/n}
\label{wnoncom}
\eeq
and hence it is necessary to specify the
order of the limits in the definition of $W(\{G\},q_s)$ \cite{w}.

Using the formula (\ref{cluster}) for $Z(G,q,v)$, one can generalize $q$ from
${\mathbb Z}_+$ not just to ${\mathbb R}_+$ but to ${\mathbb C}$ and $a$ from
its physical ferromagnetic and antiferromagnetic ranges $1 \le a \le \infty$
and $0 \le a \le 1$ to $a \in {\mathbb C}$.  A subset of the zeros of $Z$ in
the two-complex dimensional space ${\mathbb C}^2$ defined by the pair of
variables $(q,a)$ can form an accumulation set in the $n \to \infty$ limit,
denoted ${\cal B}$, which is the continuous locus of points where the free
energy is nonanalytic.  This locus is determined as the solution to a certain
$\{G\}$-dependent equation \cite{bcc,a}.  For a given value of $a$, one can
consider this locus in the $q$ plane, and we denote it as ${\cal
B}_q(\{G\},a)$.  In the special case $a=0$ (i.e., $v=-1$) where the partition
function is equal to the chromatic polynomial, the zeros in $q$ are the
chromatic zeros, and ${\cal B}_q(\{G\},a=0)$ is their continuous accumulation
set in the $n \to \infty$ limit.  In previous papers
we have given exact calculations of the chromatic polynomials and nonanalytic
loci ${\cal B}_q$ for various families of graphs; some of these are related 
earlier works are \cite{w}-\cite{bkw}-\cite{tor4} (see \cite{a} for further
refernences).  With the exact Potts partition
function for arbitrary temperature, one can study ${\cal B}_q$ for $a \ne 0$
and, for a given value of $q$, one can study the continuous accumulation set of
the zeros of $Z(G,q,v)$ in the $a$ plane; this will be denoted ${\cal
B}_a(\{G\},q)$.  It will often be convenient to consider the equivalent locus
in the $u=1/a$ plane, namely ${\cal B}_u(\{G\},q)$.  We shall sometimes write
${\cal B}_q(\{G\},a)$ simply as ${\cal B}_q$ or ${\cal B}$ 
when $\{G\}$ and $a$ are clear
from the context, and similarly with ${\cal B}_{a}$ and ${\cal B}_{u}$.  One
gains a unified understanding of the separate loci ${\cal B}_q(\{G\})$ for
fixed $a$ and ${\cal B}_a(\{G\})$ for fixed $q$ by relating these as different
slices of the locus ${\cal B}$ in the ${\mathbb C}^2$ space defined by $(q,a)$
as we shall do here.

Following the notation in \cite{w} and our other earlier works on ${\cal
B}_q(\{G\})$ for $a=0$, we denote the maximal region in the complex $q$ plane
to which one can analytically continue the function $W(\{G\},q)$ from physical
values where there is nonzero ground state entropy as $R_1$ .  The maximal
value of $q$ where ${\cal B}_q$ intersects the (positive) real axis was
labelled $q_c(\{G\})$.  Thus, region $R_1$ includes the positive real axis for
$q > q_c(\{G\})$.  Correspondingly, in our works on complex-temperature
properties of spin models, we had labelled the complex-temperature extension
(CTE) of the physical paramagnetic phase as (CTE)PM, which will simply be
denoted PM here, the extension being understood, and similarly with
ferromagnetic (FM) and antiferromagnetic (AFM); other complex-temperature
phases, having no overlap with any physical phase, were denoted $O_j$ (for
``other''), with $j$ indexing the particular phase \cite{chisq,chihc}.  
Here we shall
continue to use this notation for the respective slices of ${\cal B}$ in the
$q$ and $a$ or $u$ planes. Various special cases of $Z(G,q,v)$ applicable for
arbitrary graphs $G$ were given in \cite{a}. 

Just as one must take account of the noncommutativity in the definition of 
$W$, eq. (\ref{wnoncom}), so also one must take account of noncommutativity 
in the definition of the free energy: at certain special points $q_s$
(typically $q_s=0,1...,\chi(G)$) one has
\beq 
\lim_{n \to \infty} \lim_{q
\to q_s} Z(G,q,v)^{1/n} \ne \lim_{q \to q_s} \lim_{n \to \infty} Z(G,q,v)^{1/n}
\ .
\label{fnoncomm}
\eeq

Because of
this noncommutativity, the formal definition (\ref{ef}) is, in general,
insufficient to define the free energy $f$ at these special points $q_s$; it is
necessary to specify the order of the limits that one uses in eq.
(\ref{fnoncomm}).  We denote the two
definitions using different orders of limits as $f_{qn}$ and $f_{nq}$:
\beq
f_{nq}(\{G\},q,v) = \lim_{n \to \infty} \lim_{q \to q_s} n^{-1} \ln Z(G,q,v)
\label{fnq}
\eeq
\beq
f_{qn}(\{G\},q,v) = \lim_{q \to q_s} \lim_{n \to \infty} n^{-1} \ln Z(G,q,v) \
.
\label{fqn}
\eeq
In Ref. \cite{w} and our subsequent works on chromatic polynomials and the
above-mentioned zero-temperature antiferromagnetic limit, it was convenient to
use the ordering $W(\{G\},q_s) = \lim_{q \to q_s} \lim_{n \to \infty}
P(G,q)^{1/n}$ since this avoids certain discontinuities in $W$ that would be
present with the opposite order of limits.  In the present work on the full
temperature-dependent Potts model partition function, we shall
consider both orders of limits and comment on the differences where
appropriate.  Of course in discussions of the usual $q$-state Potts model (with
positive integer $q$), one automatically uses the definition in eq.
(\ref{zfun}) with (\ref{ham}) and no issue of orders of limits arises, as it
does in the Potts model with real $q$.
Consequences of the noncommutativity (\ref{fnoncomm}) have been discussed
before \cite{a,ta,ka}. 

As derived in \cite{a}, a general form for the Potts model partition function
for the strip graphs $G_s$ considered here, or more generally, for recursively
defined families of graphs comprised of $m$ repeated subunits (e.g. the columns
of squares of height $L_y$ vertices that are repeated $L_x=m$ times to form an
$L_x \times L_y$ strip of a regular lattice with some specified boundary
conditions), is 
\beq 
Z((G_s)_m,q,v) = \sum_{j=1}^{N_{Z,G_s,\lambda}} c_{Z,G_s,j} 
(\lambda_{Z,G_s,j}(q,v))^m
\label{zgsum}
\eeq
where $N_{Z,G_s,\lambda}$, $c_{Z,G_s,j}$, and $\lambda_{Z,G_s,j}$ depend on 
the type of recursive family $G_s$ 
(lattice structure and boundary conditions) but not on its length $m$. 
For special case of the $T=0$ antiferromagnet, the partition
function, or equivalently, the chromatic polynomial $P((G_s)_m,q)$ has the
corresponding form 
\beq
P((G_s)_m,q) = \sum_{j=1}^{N_{P,G_s,\lambda}} c_{P,G_s,j} 
(\lambda_{P,G_s,j}(q,v))^m \ .
\label{pgsum}
\eeq
For sufficiently large integer $q$ values, the coefficients can be interpreted
as multiplicities of eigenvalues of coloring matrices \cite{matmeth}, 
and the sums of these
coefficients are thus sums of dimensions of invariant subspaces of these
matrices.  For a given family $G_s$ of strip graphs we shall denote these
sums as 
\beq
C_{Z,G_s}=\sum_{j=1}^{N_{Z,G_s,\lambda}} c_{Z,G_s,j}
\label{czsum}
\eeq
and
\beq
C_{P,G_s}=\sum_{j=1}^{N_{P,G_s,\lambda}} c_{P,G_s,j} \ .
\label{cpsum}
\eeq
Here we distinguish the quantities $N_{Z,G_s,\lambda}$, $N_{P,G_s,\lambda}$; 
$c_{Z,G_s,j}$, $c_{P,G_s,j}$; $\lambda_{Z,G_s,j}$, and $\lambda_{P,G_s,j}$.  
Below, for
brevity of notation, we shall sometimes suppress the $Z$ or $P$ subscripts 
and the $s$ subscript on $G_s$ when
the meaning is clear from context and no confusion will result.

The Potts ferromagnet has a zero-temperature phase transition in the $L_x \to
\infty$ limit of the strip graphs considered here, and this has the consequence
that for cyclic and M\"obius boundary conditions, ${\cal B}$ passes through the
$T=0$ point $u=0$.  It follows that ${\cal B}$ is noncompact in the $a$ plane.
Hence, it is usually more convenient to study the slice of ${\cal B}$ in the
$u=1/a$ plane rather than the $a$ plane.  For the ferromagnet, since $a \to
\infty$ as $T \to 0$ and $Z$ diverges like $a^{e(G_s)}$ in this limit, we shall
use the reduced partition function $Z_r$ defined by
\beq
Z_r((G_s)_m,q,v)=a^{-e((G_s)_m)}Z((G_s)_m,q,v)=u^{e((G_s)_m)}Z((G_s)_m,q,v)
\label{zr}
\eeq
which has the finite limit $Z_r \to q$ as $T \to 0$.  For a general strip
graph $(G_s)_m$ of type $G_s$ and length $L_x=m$, we can write
\beqs
Z_r((G_s)_m,q,v) & = & u^{e((G_s)_m)}\sum_{j=1}^{N_{Z,G_s,\lambda}} c_{Z,G_s,j}
(\lambda_{Z,G_s,j})^m \equiv \sum_{j=1}^{N_{Z,G_s,\lambda}} c_{Z,G_s,j}
(\lambda_{Z,G_s,j,u})^m
\label{zu}
\eeqs
with
\beq
\lambda_{Z,G_s,j,u}=u^{e((G_s)_m)/m}\lambda_{Z,G_s,j} \ .
\label{lamu}
\eeq
For the Potts model partition functions of the $L_y=2$ cyclic and M\"obius 
strips of the honeycomb lattice to be discussed below, the prefactor in
eq. (\ref{lamu}) is $u^5$. 

For the following, it will be convenient to define some general functions.
First, we define the following polynomial:
\beq
D_k(q) = \frac{P(C_k,q)}{q(q-1)} = 
\sum_{s=0}^{k-2}(-1)^s {{k-1}\choose {s}} q^{k-2-s}
\label{dk}
\eeq
and $P(C_n,q)$ is the chromatic polynomial for the circuit
(cyclic) graph $C_n$ with $n$ vertices, given by $P(C_n,q) = (q-1)^n +
(q-1)(-1)^n$. 

Second, we define coefficients of degree $d$ in $q$ which are related to
Chebyshev polynomials: 
\beq
c^{(d)}=U_{2d}\Bigl (\frac{\sqrt{q}}{2} \Bigr )
\label{cd}
\eeq
where $U_n(x)$ is the Chebyshev polynomial of the second kind, defined
by 
\beq
U_n(x) = \sum_{j=0}^{[\frac{n}{2}]} (-1)^j {n-j \choose j} (2x)^{n-2j}
\label{undef}
\eeq
where in eq. (\ref{undef}) the notation $[\frac{n}{2}]$ in the upper limit on 
the summand means the integral part of $\frac{n}{2}$.  
The first few of these coefficients are
\beq
c^{(0)}=1 \ , \quad c^{(1)}=q-1 \ ,
\label{cd01}
\eeq
\beq
c^{(2)}=q^2-3q+1,
\label{cd2}
\eeq
and
\beq
c^{(3)}=q^3-5q^2+6q-1 \ .
\label{cd3}
\eeq

\section{Chromatic Polynomial for the $L_y=2$ Cyclic and M\"obius Strips of 
the Honeycomb Lattice} 

Before giving our calculation of the Potts model partition functions for these
strips, it is useful to discuss a particularly interesting special case, namely
that of the zero-temperature antiferromagnet.  For the open strip of $m$ 
bricks, which will be denoted $S_m$ for short, 
\beq
P(hc,2 \times m, FBC_y,FBC_x,q)=q(q-1)(D_6)^m=
q(q-1)(q^4-5q^3+10q^2-10q+5)^m
\label{popen}
\eeq
so that in the $m \to \infty$ limit, 
\beq
W(hc,2 \times \infty, FBC_y,FBC_x,q)=(q^4-5q^3+10q^2-10q+5)^{1/4} \ .
\label{wopen}
\eeq
Since this has only discrete branch point singularities at the zeros of $D_6$,
the continuous locus ${\cal B}=\emptyset$.  

The chromatic polynomials for width $L_y=2$ cyclic and M\"obius strips of the
honeycomb lattice can be obtained as special cases of the calculations in
\cite{pg} for homeomorphic expansions of cyclic and M\"obius strips of the
square lattice.  These homeomorphic expansions involved the addition of $k-2$
degree-2 vertices on the upper and lower horizontal edges of the strips,
thereby making cyclic or M\"obius strips of $p$-gons with $p=2k$ such that two
successive $p$-gons intersected on one common edge.  Here, the degree $\Delta$
(i.e., coordination number) of a vertex in a graph $G$ is defined as the 
number of edges (bonds) that connect to it.  The case $k=3$, i.e.,
$p=6$, was thus the width 2 cyclic or M\"obius strip of the honeycomb
lattice, depending on the longitudinal boundary conditions.  We
denote these strips as $L_m=hc,2 \times m,FBC_y,PBC_x$ and 
$ML_m=hc,2 \times m,FBC_y,TPBC_x$ for short.  For quantities that are
independent of the value of $m$ in $L_m$ or $ML_m$, we shall use the notation
$L$ and $ML$.  The graphs $L_m$ and $ML_m$ have $n=4m$ vertices. 
For $m$ large enough to avoid degenerate cases (i.e. $m \ge 2$ for the cyclic
strip and $m \ge 1$ for the M\"obius strip),
the chromatic number for the $L_y \times m$ strips of
the honeycomb lattice with cyclic boundary conditions is
\beq
\chi(hc, L_y \times m,FBC_y,PBC_x) = 2 
\label{chi_hc_cyclic}
\eeq
while for the M\"obius strips we have 
\beq
\chi(hc, L_y \times m,FBC_y,TPBC_x) = 3
\label{chi_hc_mb}
\eeq
independent of $m$ in the indicated ranges. 

The results for the chromatic polynomials (in our current notation) are
\cite{pg} 
\beq
P(hc, 2 \times m, FBC_y,PBC_x,q)=\sum_{j=1}^4 c_{P,L,j}
(\lambda_{P,L,j})^m
\label{phc2}
\eeq
and
\beq
P(hc, 2 \times m, FBC_y,TPBC_x,q)=\sum_{j=1}^4 c_{P,ML,j}
(\lambda_{P,L,j})^m
\label{phc2mb}
\eeq
where, as indicated, the $\lambda_{P,G,j}$'s 
are the same for cyclic and M\"obius
strips \cite{pg,pm}:
\beq
\lambda_{P,L,1}=1
\label{lamhc2_1}
\eeq
\beq
\lambda_{P,L,2}=D_4+D_3=(q-1)^2
\label{lamhc2_2}
\eeq
\beq
\lambda_{P,L,3}=D_4-D_3=q^2-4q+5
\label{lamhc2_3}
\eeq
\beq
\lambda_{P,L,4}=D_6=q^4-5q^3+10q^2-10q+5 \ . 
\label{lamhc2_4}
\eeq

The corresponding coefficients are 
\beq
c_{P,L,1}=c^{(2)}
\label{chc2_1}
\eeq
\beq
c_{P,L,2}=c_{P,L,3}=c^{(1)}
\label{chc2_23}
\eeq
\beq
c_{P,L,4}=c^{(0)}=1
\label{chc2_4}
\eeq
\beq
c_{P,ML,1}=-c^{(0)}=-1
\label{chc2mb_1}
\eeq
\beq
c_{P,ML,2}=-c^{(1)}
\label{chc2mb_2}
\eeq
\beq
c_{P,ML,3}=c^{(1)}
\label{chc2mb_3}
\eeq
\beq
c_{P,ML,4}=c^{(0)}=1 \ .
\label{chc2mb_4}
\eeq
The equality of the coefficients (here, $c_{P,L,4}$ and $c_{P,ML,4}$) of the
$\lambda_{G,j}$ that is dominant in region $R_1$, independent of longitudinal
boundary conditions, is a general result \cite{pm,bcc}.

\begin{figure}[hbtp]
\centering
\leavevmode
\epsfxsize=2.5in
\begin{center}
\leavevmode
\epsffile{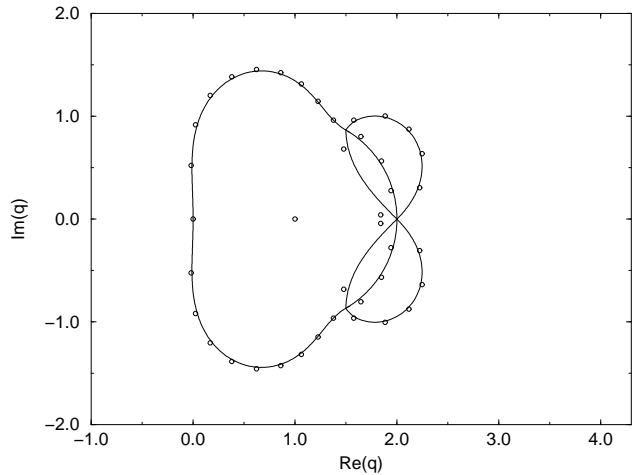}
\end{center}
\caption{\footnotesize{Locus ${\cal B}_q$ for the $m \to \infty$ limit of
the cyclic $L_y=2$ strip $\{L\}$ of the honeycomb lattice with $a=0$. 
Zeros of $Z(L_m,q,v=-1)=P(L_m,q)$ for $m=20$, i.e., $n=80$ vertices, are 
shown for comparison.}}
\label{hpxy2}
\end{figure}

The singular locus ${\cal B}$ was given as Fig. 1(a) in \cite{pg}.  For our
current generalization to nonzero temperature, it will be necessary to refer to
details of this locus, so we show it as Fig. \ref{hpxy2}.  As is evident in
this figure, the locus (boundary) ${\cal B}$ separates the $q$ plane into six
regions: (i) $R_1$ containing the real intervals $q > q_c=2$ and $q < 0$ and
extending outward to complex infinity, (ii) an innermost region $R_2$
containing the real interval $0 < q < q_c$, and two complex conjugate pairs of
regions, (iii) $R_3,R_3^*$ to the upper and lower right of $q_c$, and (iv)
$R_4,R_4^*$ to the upper and lower left of $q_c$.  
The dominant terms in these regions are (i)
$\lambda_{P,L,4}$ in $R_1$, (ii) $\lambda_{P,L,3}$ in $R_2$, (iii)
$\lambda_{P,L,2}$ in $R_3,R_3^*$, and (iv) $\lambda_{P,L,1}$ in $R_4,R_4^*$.
The point $q_c$ is a multiple point where six curves, forming three branches of
${\cal B}$, cross.  This corresponds to the fact that at $q_c$ there is a
degeneracy in magnitude of all of the $\lambda_{P,L,j}$'s.  The angles at which
the various boundary curves emanates from $q_c$ are: (a) $\pi/4$ for the curve
separating $R_1$ and $R_3$, (b) $\pi/2$ for the curve separating $R_3$ from
$R_4$, and (c) $3\pi/4$ for the curve separating $R_4$ from $R_2$, and so forth
for the complex-conjugate curves.  The density of chromatic zeros is small on
the boundary separating $R_2$ and $R_4$, and its complex conjugate boundary.
The locus ${\cal B}$ has support for $Re(q)
< 0$ as well as $Re(q) \ge 0$.  Further, some chromatic zeros for finite-$m$
strips and a portion of the locus ${\cal B}$ for the $m \to \infty$ limit both
lie within the circle $|q-1|=1$.  This is also true for the $L_y=3$ cyclic and
M\"obius strips of the honeycomb lattice (see below).  In this context, we
recall that the conjecture later made in \cite{s4,t} that chromatic zeros and,
in the limit of infinite-length, the locus ${\cal B}$, do not lie in the
interior of the circle $|q-1|$ was stated to be specific to strips of the 
square and triangular lattice.

The $W$ function in these regions is given by 
\beq
W=(D_6)^{1/4} \quad {\rm for} \ \ q \in R_1
\label{wr1hcly2cyc}
\eeq
\beq
|W|=|q^2-4q+5|^{1/4} \quad {\rm for} \ \ q \in R_2
\label{wr2hcly2cyc}
\eeq
\beq
|W|=|q-1|^{1/2} \quad {\rm for} \ \ q \in R_3,R_3^*
\label{wr3hcly2cyc}
\eeq
\beq
|W|=1 \quad {\rm for} \ \ q \in R_4,R_4^* \ .
\label{wr4hcly2cyc}
\eeq
Recall that for regions other than $R_1$, only the magnitude of $W$ can be 
determined unambiguously \cite{w}.  
Note that $W(q=2)=1$.  We next generalize this study
to the case of finite temperature for both the antiferromagnet and the
ferromagnet.

\section{Potts Model Partition Function for Open $L_y=2$ Strip of the Honeycomb
Lattice}

By using an iterative application of the deletion-contraction theorem
for Tutte polynomials and converting the result to $Z$, or by using a
transfer matrix method, one can calculate the
partition function for the open, cyclic, and M\"obius strips of the honeycomb
lattice of width $L_y$ and arbitrary length $m$ bricks, 
which we shall again denote as $S_m$, $L_m$, and $ML_m$.
We have used both methods.  These calculations are a natural extension of our
earlier calculations of Potts model partition functions for finite-width
lattice strips of arbitrarily great length \cite{bcc}, \cite{a}-\cite{ka}. 
We begin with the open strip.  The results are conveniently expressed 
in terms of a generating function
\beq
\Gamma_Z(S_m,q,v)=\sum_{m=0}^\infty Z(S_m,q,v)z^m \ .
\label{gammas}
\eeq
We find
\beq
\Gamma_Z(S_m,q,v)=\frac{A_{S,0}+A_{S,1}z}{
(1-\lambda_{Z,S,1}z)(1-\lambda_{Z,S,2}z)}
\label{gammand}
\eeq
where
\beq
A_{S,0}=q(q+v)
\label{as0hc}
\eeq
\beq
A_{S,1}=-q^2v^4(1+v)
\label{as1hc}
\eeq
\beq
\lambda_{Z,S,(1,2)}=\frac{1}{2}(T_S \pm \sqrt{R_S})
\label{lams12}
\eeq
with
\beq
T_S=v^5+6v^4+10qv^3+10q^2v^2+5q^3v+q^4
\label{ts}
\eeq
\beqs
R_S & = & 4qv^8-4q^2v^7-6q^3v^6-2q^4v^5+v^{10}+32v^8+8v^9+104v^7q \cr\cr
& & +196v^6q^2+244v^5q^3+208v^4q^4+q^8+10q^7v+45q^6v^2+120q^5v^3 \ .
\label{rs}
\eeqs
The equivalent Tutte polynomial $T(S_m,x,y)$ is given in the appendix. 

\begin{figure}[hbtp]
\centering
\leavevmode
\epsfxsize=2.5in
\begin{center}
\leavevmode
\epsffile{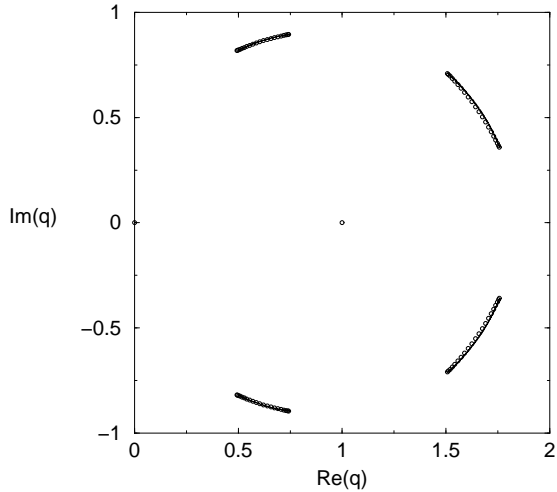}
\end{center}
\caption{\footnotesize{Locus ${\cal B}_q$ for the $n \to \infty$ limit of
the $L_y=2$ honeycomb strip $\{S\}$ with free longitudinal boundary
conditions, for $a=v+1=0.1$. Zeros of $Z(S_m,q,v=-0.9)$ for $m=21$, i.e.,
$n=86$ vertices, are shown for comparison.}}
\label{hy2a0p1}
\end{figure}

We have studied the locus ${\cal B}$ in the ${\mathbb C}^2$ space defined by
the variables $(q,v)$ (or equivalently, $(q,a)$).   For $a=0$, $Z$ 
reduces to the chromatic polynomial (\ref{popen}), and the resultant $W$
function, given in eq. (\ref{wopen}) has only discrete branch point 
singularities at the zeros of $D_6$ (namely, $q \simeq 0.690983 \pm 
0.9510565i$ and $1.809017 \pm 0.587785i$), and consequently 
${\cal B}=\emptyset$. As $a$ initially increases from 0, ${\cal B}_q$ 
consists of arcs that spread out from these four points.  An illustrative 
case, $a=0.1$, is shown in Fig. \ref{hy2a0p1}. As $a \to 1$, these arcs shrink
in toward the origin, $q=0$. 

\begin{figure}[hbtp]
\centering
\leavevmode
\epsfxsize=2.5in
\begin{center}
\leavevmode
\epsffile{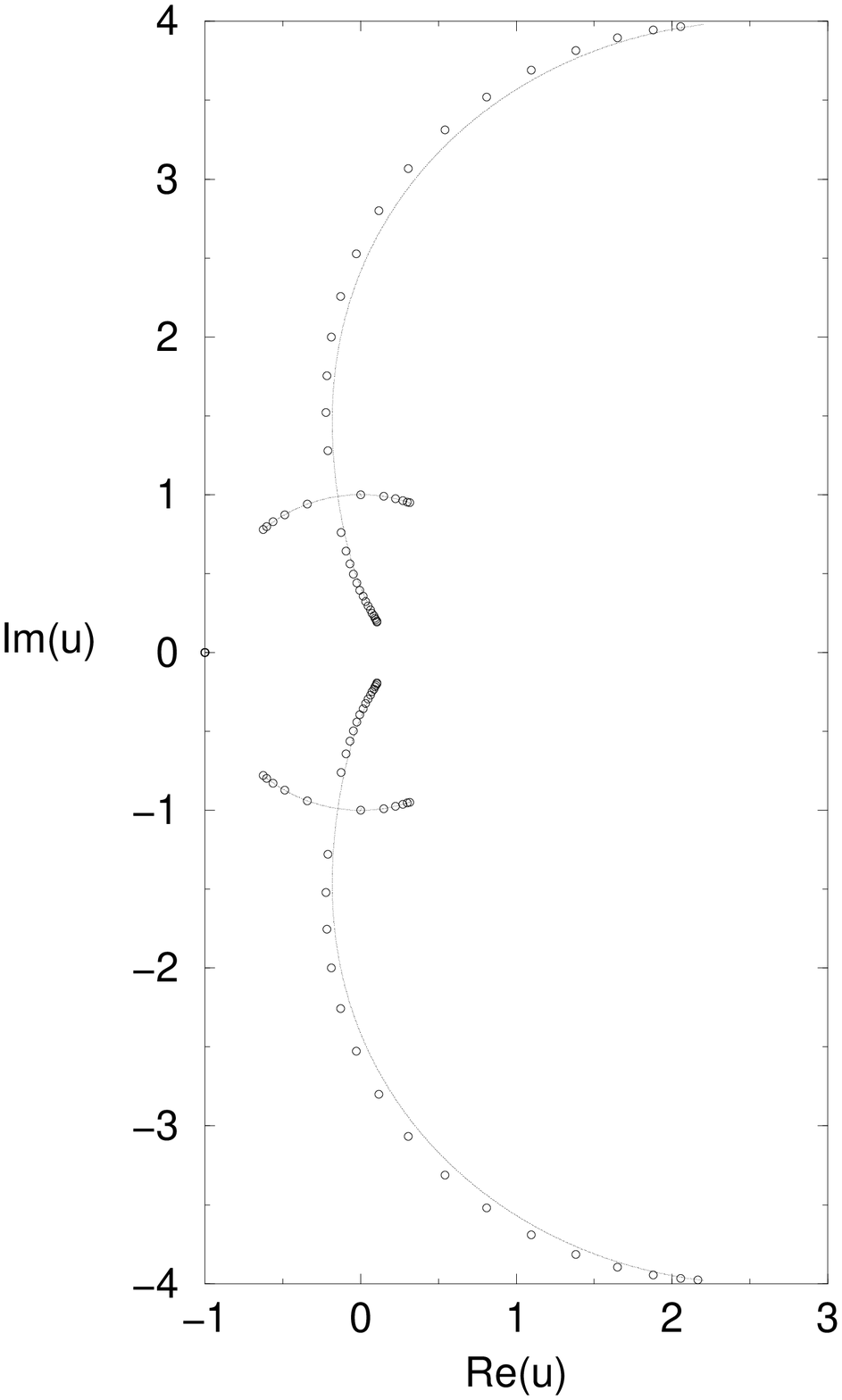}
\end{center}
\caption{\footnotesize{Locus ${\cal B}_u$ for the $m \to \infty$ limit of the
$L_y=2$ strip of the honeycomb lattice with free longitudinal boundary
conditions, $\{S\}$ with $q=2$. Zeros of $Z(S_m,q=2,v)$ in $u$ for $m=21$ (so
that $Z$ is a polynomial of degree 106 in $v$ and hence, up to an overall
factor, in $u$) are shown for comparison.}}
\label{hy2q2}
\end{figure}

\begin{figure}[hbtp]
\centering
\leavevmode
\epsfxsize=2.5in
\begin{center}
\leavevmode
\epsffile{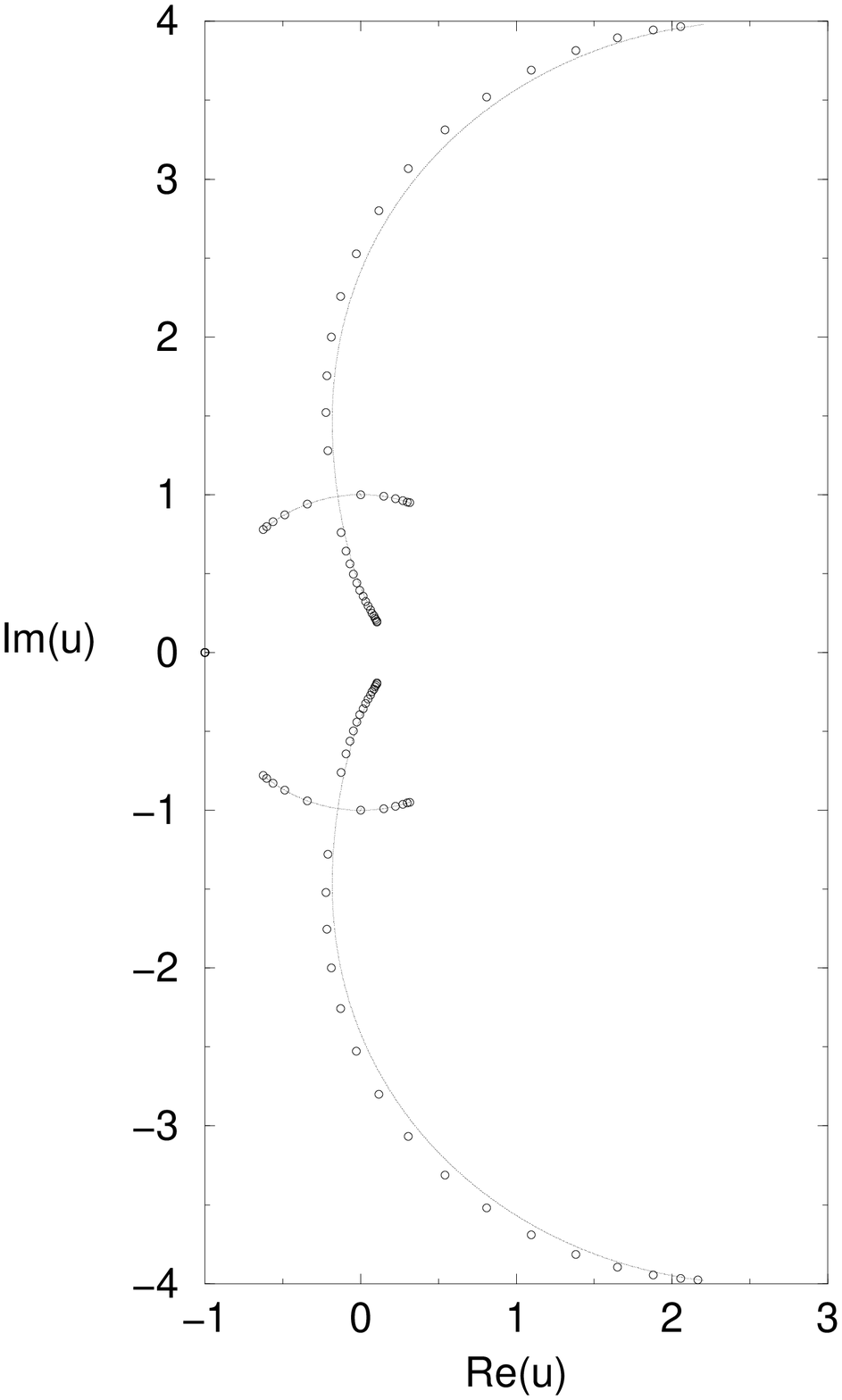}
\end{center}
\caption{\footnotesize{Locus ${\cal B}_u$: same as in Fig. \ref{hy2q2} 
for $q=3$.}}
\label{hy2q3}
\end{figure}

In Figs. \ref{hy2q2} and \ref{hy2q3} we show ${\cal B}_u$ for $q=2$ and 3,
respectively.  The positive real $u$ axis, and the region reached by
analytic continuation from it, form the (CTE)PM phase.  Since the
infinite-length, finite-width strips are quasi-one-dimensional systems and the
spin-spin couplings are short-range (specifically, nearest-neighbor), an
elementary application of a Peierls argument shows that there is no long-range
order and thus no broken-symmetry (FM or AFM) phases at any finite
temperature.  Related to this, 
the singular locus ${\cal B}_u$ does not cross the positive real $u$ axis.

\section{Thermodynamic Properties of the Potts Model on the $2 \times \infty$
Strip of the Honeycomb Lattice}

We next discuss the physical thermodynamic properties of the $q$-state Potts
model on the $2 \times \infty$ strip of the honeycomb lattice.  These are
independent of the longitudinal boundary conditions.  The reduced free 
energy per site is given (with the $f_{nq}$ definition) by 
\beq
f = \frac{1}{4}\ln \lambda_{Z,S,1} \ .
\label{fhc}
\eeq
This is analytic for all finite temperature for both signs of the spin-spin
coupling (ferromagnetic and antiferromagnetic).  This analyticity property is 
equivalent to 
the above-mentioned fact that the singular locus ${\cal B}_u$ does not cross 
the positive real $u$ axis.  It is straightforward to
calculate from this the internal energy $U$ and specific heat $C$ per site.  
These have the high-temperature expansions
\beq
U=-\frac{5J}{4q}\biggl [ 1 + \Big ( \frac{q-1}{q} \Bigr ) K + O(K^2) \biggr ]
\label{uhigh}
\eeq
\beq
C=\frac{5k_B(q-1)K^2}{4q^2} \biggl [ 1 + \Bigl ( \frac{q-2}{q} \Bigr ) K 
+ O(K^2) \biggr ] \ .
\label{chigh}
\eeq

\begin{figure}[hbtp]
\centering
\leavevmode
\epsfxsize=2.5in
\begin{center}
\leavevmode
\epsffile{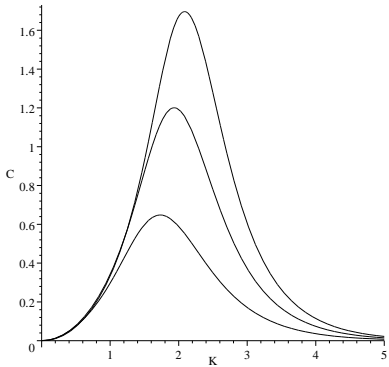}
\end{center}
\vspace*{-3cm}
\caption{\footnotesize{Specific heat $C/k_B$ for the $q$-state Potts
ferromagnet on the $2 \times \infty$ strip of the honeycomb lattice as a
function of $K=J/(k_BT)$.  From bottom to top the curves are for $q=2,3,4$.}}
\label{cfmhcstrip}
\end{figure}

In Fig. \ref{cfmhcstrip} we show a plot of the specific heat for the
ferromagnetic case.  One sees that as $q$ increases, the value of $K$ at 
which the maximum in the specific heat occurs increases; i.e. this maximum
occurs at lower temperature.  This is a consequence of the fact that the
spins are ``floppier'' for larger $q$, and hence one must cool the system to
lower temperatures in order to achieve short-range ordering (as noted above,
there is, of course, no long-range ordering). 
Further, as $q$ increases, the height of the maximum increases.
This can be interpreted as a result of the fact that as one makes $q$ larger,
one increases the effective number of degrees of freedom per site, and it is
the onset of short-range ordering of these degrees of freedom that produces 
the maximum in the specific heat.

\begin{figure}[hbtp]
\centering
\leavevmode
\epsfxsize=2.5in
\begin{center}
\leavevmode
\epsffile{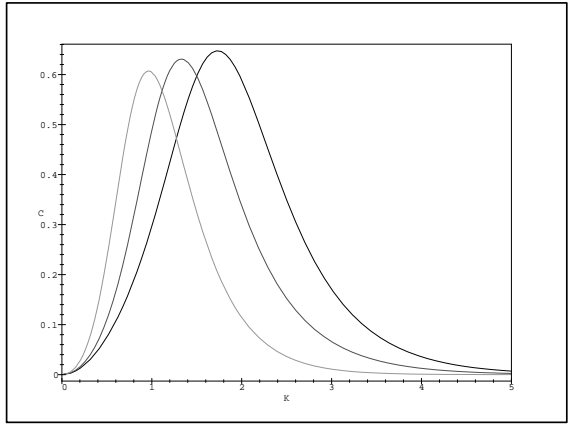}
\end{center}
\vspace*{-1cm}
\caption{\footnotesize{Comparison of the specific heat $C/k_B$ (vertical axis)
for the $q$-state Potts ferromagnet on the $2 \times \infty$ lattice strips for
the Ising case, $q=2$.  In order of increasing position $K_{max}$ of the maxima
in these curves, they apply for the (a) triangular, (b) square, and (c)
honeycomb.}}
\label{c_3lat_q2}
\end{figure}

It is also of interest to compare our exact calculations of the specific heat
for the $2 \times \infty$ strip of the honeycomb lattice with our previous
calculations for the $2 \times \infty$ strips of the square and triangular
lattices.  In Fig. \ref{c_3lat_q2} we show this comparison
for the Ising case $q=2$.  One sees that the value of $K$ 
at which the maximum occurs increases as one goes from triangular to
square to honeycomb lattices.  This is a consequence of the fact that the
short-range ordering in the spins is strengthened as one increases the
effective coordination number (vertex degree) of the lattice strip. The cyclic
$L_y=2$ strips of the square and triangular lattices are
$\Delta$-regular, i.e., each vertex has the same degree, namely 
$\Delta=3$ and $\Delta=4$, respectively.  This is also true of the
infinite-length limit of the open $L_y=2$ strips of the square and triangular
lattices.  For the $L_y=2$ strip of the
honeycomb lattice, and more generally, for a graph $G$ and its $n \to \infty$
limit $\{G\}$, it is useful to define the effective coordination number 
(degree) as 
\beq
\Delta_{eff}(\{G\}) = \lim_{n(G) \to \infty} \frac{2e(G)}{n(G)} \ . 
\label{delta_eff}
\eeq
For the infinite-length limit of the $L_y=2$ strip of the
honeycomb lattice, half of the vertices have degree 2 and half have degree 3,
so that the effective coordination number is 2.5.  Thus, since one increases 
the (effective) coordination number from 2.5 to 3 to 4 going from the honeycomb
to square to triangular $L_y=2$ strips, one can achieve short-range order at a
progressively higher temperature (smaller value of $K$).   The same qualitative
comparison applies for $q=3,4$, etc. 

Since the Potts ferromagnet has a $T=0$ critical point on infinite-length,
finite-width strips, it is worthwhile to examine the critical singularities.  
The specific heat has the low-temperature behavior
\beq
C = 6k_BK^2(q-1)e^{-2K}\biggl [1 + \frac{3}{2}(q+2)e^{-K} + O(e^{-2K}) 
\biggr ] \quad {\rm as} \ \ K \to \infty \ .
\label{cfmlow}
\eeq 
Thus, as is typical of systems at their lower critical dimensionality and
was true of the other finite-width, infinite-length strips for which we have
performed exact calculations \cite{a,ta,ka}, the specific heat exhibits an
essential zero at $T=0$ as a function of temperature. 
In general, at a critical
point the ratio $\rho$ of the largest subdominant to the dominant
$\lambda_j$'s determines the asymptotic decay of the connected spin-spin
correlation function and hence the correlation length
\beq
\xi = -\frac{1}{\ln \rho} \ .
\label{xi}
\eeq
Since $\lambda_{Z,L,5}$ and $\lambda_{Z,L,3}$ are the dominant and leading
subdominant $\lambda_{Z,G,j}$'s, respectively, we have
\beq
\rho_{FM}=\frac{\lambda_{Z,L,3}}{\lambda_{Z,L,5}}
\label{rho}
\eeq
and hence for the ferromagnetic zero-temperature critical point we find that
the correlation length diverges, as $T \to 0$, as
\beq
\xi_{FM} \sim q^{-1}e^{2K} + O(e^K) \ , \quad {\rm as} \quad K \to \infty
\ .
\label{xit}
\eeq 
Again, this is an essential singularity (divergence) as a function of
temperature, as is generic for systems at their lower critical dimensionality. 
The detailed form of the exponent is similar to the behavior 
$\xi \sim q^{-1}e^{2K}$ of the correlation
length at the $T=0$ critical point of the Potts
ferromagnet on the $L_y=2$ infinite-length strip of the square lattice
\cite{a}.  While the correlation lengths for other 1D or quasi-1D systems also
exhibit divergences with essential singularities, the detailed form of the
exponent (coefficient of $K$) varies.  Thus, for example, 
$\xi_{FM} \sim q^{-1}e^{K}$ as $K \to \infty$
for the 1D Potts ferromagnet, while $\xi_{FM}
\sim q^{-1}e^{3K}$ and $\xi_{FM} \sim q^{-1}e^{4K}$ as $K \to \infty$ for the
Potts ferromagnet on the $L_y=2$ strips of the triangular lattice \cite{ta} and
on the square lattice with next-nearest-neighbor spin-spin couplings,
respectively.

\begin{figure}[hbtp]
\centering
\leavevmode
\epsfxsize=2.5in
\begin{center}
\leavevmode
\epsffile{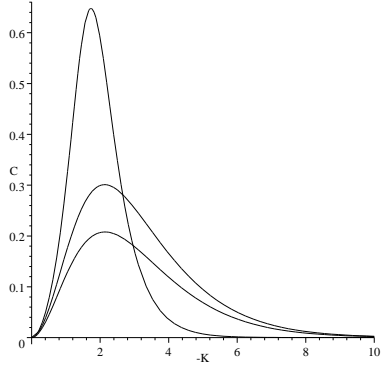}
\end{center}
\vspace*{-3cm}
\caption{\footnotesize{Specific heat $C/k_B$ for the $q$-state Potts
antiferromagnet on the $2 \times \infty$ strip of the honeycomb lattice as a
function of $-K=-J/(k_BT) > 0$. From top to bottom, the curves are for
$q=2,3,4$.}}
\label{cafmhcstrip}
\end{figure}

Except for the $q=2$ Ising case, where, as is true on any bipartite lattice,
there is an elementary equivalence between the ferromagnetic and
antiferromagnetic signs of the spin-spin coupling $J$, the $q$-state Potts
antiferromagnet on the $2 \times \infty$ strip of the honeycomb lattice is not
critical at $T=0$.  This is reflected in the fact that ${\cal B}_q$ only
crosses the positive real axis at $q=2$.  For $q \ne 2$, the low-temperature
behavior of the specific heat is given by $C \sim const. \times a$ as $a \to
0$, while for $q=2$, $C \sim const. \times a^2$ as $a \to 0$, as follows from
the equivalence and eq. (\ref{cfmlow}).  A plot of the specific heat for this
antiferromagnetic case is given in Fig. \ref{cafmhcstrip}.

\section{Potts Model Partition Function for Cyclic and M\"obius $L_y=2$ 
Strips of the Honeycomb Lattice}

\subsection{Results for $Z$}

For the cyclic and M\"obius $L_y=2$ strips of the honeycomb lattice of length
$m$ bricks $(G_s)_m=L_m, \ ML_m$, using the same methods as we did for the open
strip discussed above, we calculate the partition functions
\beq
Z((G_s)_m,q,v) = \sum_{j=1}^6 c_{Z,G_s,j}(\lambda_{Z,G_s,j}(q,v))^m \quad 
{\rm for} \ \ G_s = L, ML
\label{zgsuml}
\eeq
where
\beq
\lambda_{Z,L,j}=\lambda_{Z,ML,j} \ , j=1,...,6
\label{lamlml}
\eeq
\beq
\lambda_{Z,L,1}=v^4
\label{lam1}
\eeq
\beq
\lambda_{Z,L,2}=v^2(q+v)^2
\label{lam2}
\eeq
\beq
\lambda_{Z,L,(3,4)}=\frac{v^2}{2}\Bigl [ q^2+4qv+6v^2+v^3 \pm \sqrt{R_{34}} \ 
\Bigr ]
\label{lam34}
\eeq
with
\beq
R_{34}=40qv^3+24q^2v^2+8q^3v-2q^2v^3+8v^5+32v^4+q^4+v^6
\label{r34}
\eeq
and
\beq
\lambda_{Z,L,5}=\lambda_{Z,S,1} \ , \quad \lambda_{Z,L,6}=\lambda_{Z,S,2} \ .
\label{lam56}
\eeq

For the cyclic strip the coefficients in eq. (\ref{zgsuml}) are
\beq
c_{Z,L,1}=c^{(2)}=q^2-3q+1
\label{c1cyc}
\eeq
\beq
c_{Z,L,j}=c^{(1)}=q-1 \quad {\rm for} \quad j=2,3,4
\label{c234cyc}
\eeq
\beq
c_{Z,L,j}=c^{(0)}=1 \quad {\rm for} \quad j=5,6 \ .
\label{c56cyc}
\eeq
For the M\"obius strip, 
\beq
c_{Z,ML,1}=-c^{(0)}=-1
\label{c1zmb}
\eeq
\beq
c_{Z,ML,2}=-c^{(1)}
\label{cz2mb}
\eeq
\beq
c_{Z,ML,j}=c^{(1)} \quad {\rm for} \quad j=3,4
\label{c34mb}
\eeq
\beq
c_{Z,ML,j}=1  \quad {\rm for} \quad j=5,6 \ .
\label{c56mb}
\eeq
These are the same respective coefficient functions that we found for 
$q$-state Potts model partition function for the width $L_y=2$ cyclic and
M\"obius strips of the square lattice and for the cyclic strip of the 
triangular lattice 
(for the M\"obius strip of the triangular lattice, the coefficients are
algebraic rather than polynomial functions of $q$) \cite{bcc,a,ta}. 
The respective sums of the coefficients are 
\beq
C_{Z,L}=q^2 
\label{czsumcyc}
\eeq
and
\beq
C_{Z,ML}=q  \ . 
\label{czsummob}
\eeq

In the $T=0$ (i.e., $v=-1$) case for the Potts antiferromagnet, where the
partition function reduces to the chromatic polynomial, two of the six
$\lambda_{Z,L,j}$'s vanish and the others reduce to those for the $L_y=2$
cyclic/M\"obius strip given in \cite{pg} and above:
\beq
\lambda_{Z,L,1} = \lambda_{P,L,1} \quad {\rm for} \ \ v=-1
\label{lam1a0}
\eeq
\beq
\lambda_{Z,L,2} =\lambda_{P,L,2}=(q-1)^2 \quad {\rm for} \ \ v=-1
\label{lam2a0}
\eeq
\beq
\lambda_{Z,L,3}=\lambda_{P,L,3}=q^2-4q+5 \quad {\rm for} \ \ v=-1
\label{lam3a0}
\eeq
\beq
\lambda_{Z,L,j}=0  \quad {\rm for} \ \ v=-1 \ \ {\rm and} \ \ j=4,6
\label{lam46a0}
\eeq
\beq
\lambda_{Z,L,5}=\lambda_{P,L,4}=D_6=q^4-5q^3+10q^2-10q+5
\quad {\rm for} \ \ v=-1 \ .
\label{lam5a0}
\eeq
Here the sums of the corresponding coefficients are
\beq
C_{P,L}=q(q-1)
\label{cpsumcyc}
\eeq
and
\beq
C_{P,ML}=0  \ . 
\label{cpsummob}
\eeq

For infinite temperature, $v=0$, we have
\beq
\lambda_{Z,L,5}=q^4  \quad {\rm for} \ \ v=0
\label{lam5v0}
\eeq
\beq
\lambda_{Z,L,j}=0 \quad {\rm for} \ \ v=0 \ \ {\rm and} \ \ j=1,2,3,4,6
\label{lamne5v0}
\eeq
so that the partition function reduces to 
\beq
Z(L_m,q,0)=Z(ML_m,q,0)=(\lambda_{Z,L,5})^m=q^{4m}=q^n \ .
\label{zv0}
\eeq

 From our exact calculations, it follows that in the full ${\mathbb C}^2$
space spanned by $(q,v)$, 
\beq
{\cal B}(\{L\}) = {\cal B}(\{ML\}) \ .
\label{bcycmob}
\eeq
This is the same result that we obtained for the analogous strips of other
lattices \cite{bcc,a,ta,ka} and is in accord with the conclusion that the
singular locus is the same for an infinite-length finite-width strip graph for
given transverse boundary conditions, independent of whether the longitudinal 
boundary condition is cyclic or M\"obius.  For ${\cal B}_q$ we have proved this
in \cite{tor4} by the use of crossing-subgraph strips, which subsume cyclic and
M\"obius strips as special cases (see further below). 
(Of course, the locus ${\cal B}$ is
different if one uses open, as contrasted with cyclic/M\"obius, longitudinal
boundary conditions.) 
Owing to the equality (\ref{bcycmob}), we shall henceforth, for
brevity of notation, refer to both ${\cal B}(\{L\})$ and ${\cal B}(\{ML\})$ as
${\cal B}(\{L\})$ and similarly for specific points on ${\cal B}$, such as
$q_c(\{L\})=q_c(\{ML\})$, etc.

\subsection{${\cal B}$ in the $q$ Plane}

\begin{figure}[hbtp]
\centering
\leavevmode
\epsfxsize=2.5in
\begin{center}
\leavevmode
\epsffile{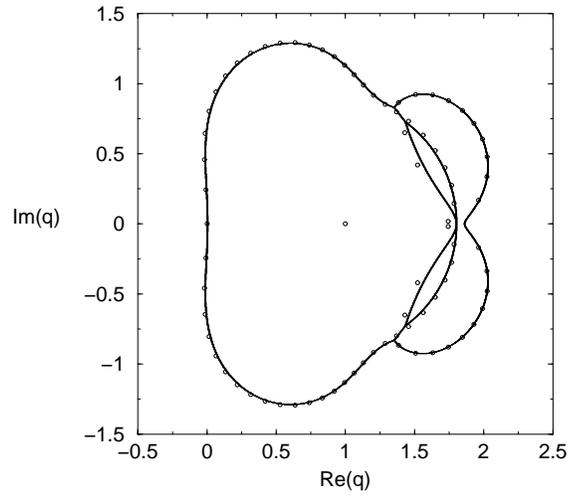}
\end{center}
\caption{\footnotesize{Locus ${\cal B}_q(\{L\})$ for the Potts antiferromagnet
on the $m \to \infty$ limit of the cyclic or M\"obius $L_y=2$ strip of the
honeycomb lattice for $a=0.1$. For comparison, chromatic zeros are shown for
the cyclic strip with $m=20$ (i.e., $n=80$).}}
\label{hpxy2a0p1}
\end{figure}

\begin{figure}[hbtp]
\centering
\leavevmode
\epsfxsize=2.5in
\begin{center}
\leavevmode
\epsffile{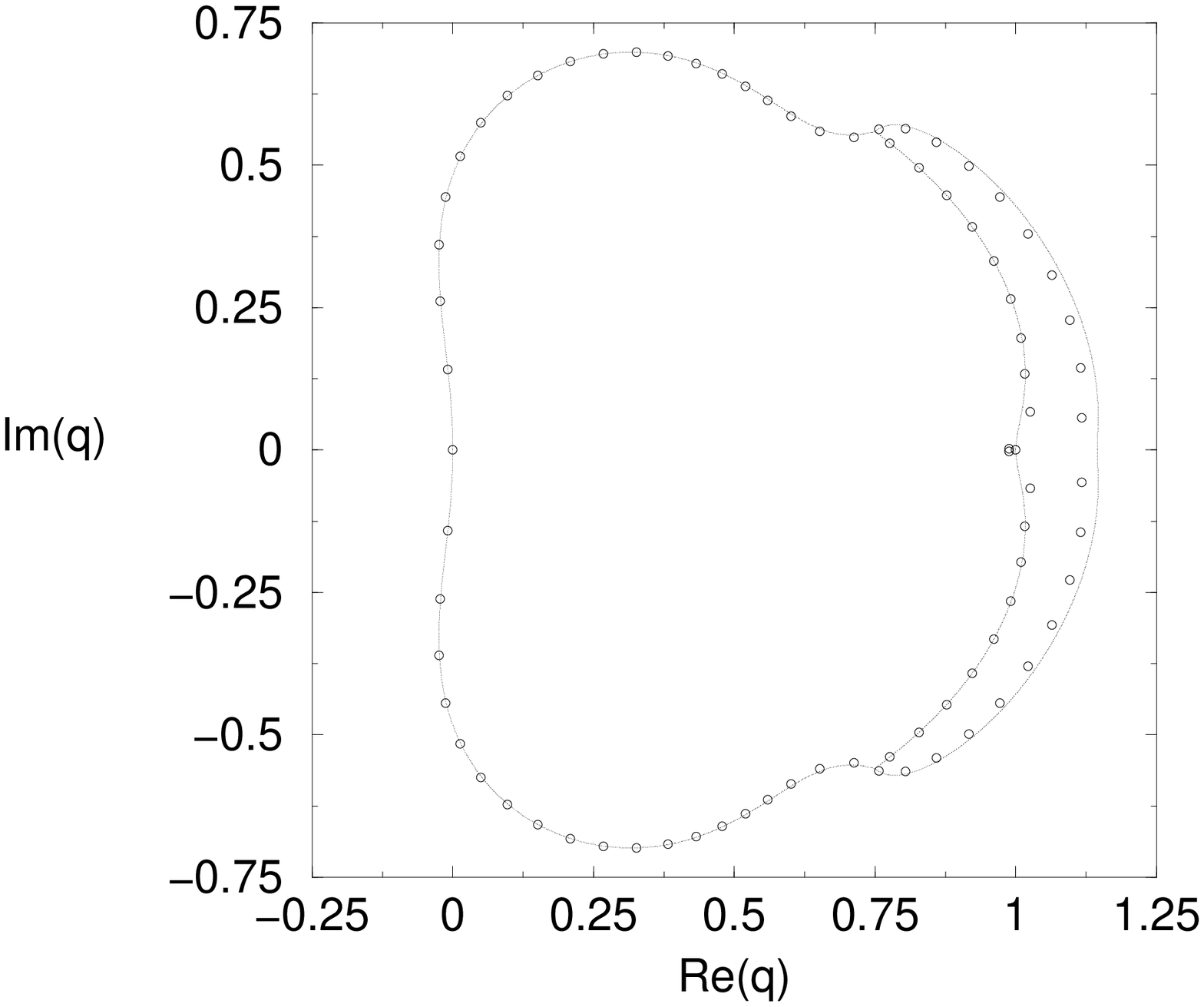}
\end{center}
\caption{\footnotesize{Same as Fig. \ref{hpxy2a0p1}, for $a=0.5$.}}
\label{hpxy2a0p5}
\end{figure}

\begin{figure}[hbtp]
\centering
\leavevmode
\epsfxsize=2.5in
\begin{center}
\leavevmode
\epsffile{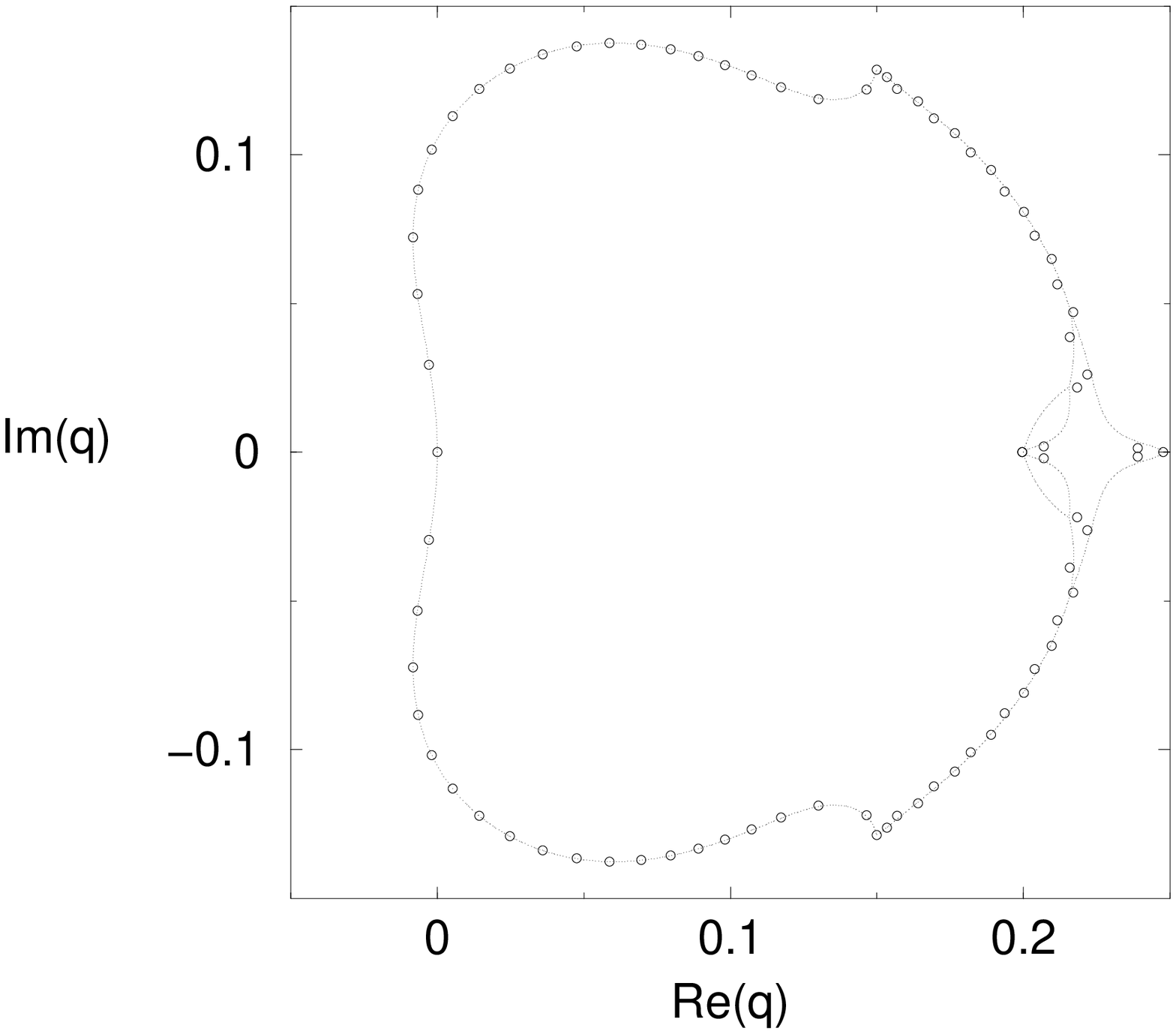}
\end{center}
\caption{\footnotesize{Same as Fig. \ref{hpxy2a0p1}, for $a=0.9$.}}
\label{hpxy2a0p9}
\end{figure}

The locus ${\cal B}$ exhibits a number of qualitative features similar to those
that we have discussed in earlier works \cite{bcc,a,ta,ka}.  Let us first
consider the slice of this locus in the $q$ plane, ${\cal B}_q$ as a function
of $v$, or equivalently, $a=v+1$.  As $a$ increases from 0, the single crossing
of ${\cal B}_q$ which had been at $q=2$ for $a=0$ splits into two crossings. At
the larger one, the boundary is determined by the equation of the degeneracy of
magnitudes of 
leading terms $|\lambda_{Z,L,5}|=|\lambda_{Z,L,2}|$, which yields the
value of $q_c$ as the real solution of the cubic equation
$q^3+3vq^2+v^2(2-v)q+2v^3=0$. The second crossing is determined by the equation
of degeneracy of magnitudes of $|\lambda_{Z,L,j}|$ for $j=1,2,4,6$ and 
occurs, for $0 \le a \le 1$, at
\beq
q_{inner}=-2v=2(1-a) \ . 
\label{qcinner}
\eeq 
or equivalently, $a=-(q_{inner}-2)/2$. 
Parenthetically, we note that this is the value for $q_c$, as a function
of $a$, for the infinite-length limit of the circuit graph \cite{is1d,a}.  
In the interval $0 
< q < q_{inner}$, $\lambda_{Z,L,3}$ is dominant, and, as was the case at
$a=0$, the crossing of ${\cal B}_q$ at $q=0$ is determined by the equation of
degeneracy of leading terms $|\lambda_{Z,L,5}|=|\lambda_{Z,L,3}|$.  For
sufficiently small $a$ near 0, there is also a pair of complex-conjugate
regions where $\lambda_{Z,L,1}$ is dominant. 
The outermost portion of ${\cal B}_q$ crosses the real
axis at $q=q_c$ and $q=0$; outside of this is the region $R_1$, including the
real intervals $q > q_c$ and $q < 0$.  As illustrations, we show ${\cal B}_q$
in Figs. \ref{hpxy2a0p1}-\ref{hpxy2a0p9} for $a=0.1$, 0.5, and 0.9,
respectively.  For the case $a=0.9$ we note that (a) the boundary includes 
cusp-like structures at $q_c$ and $q_{inner}$, and (b) in the pair of
small complex-conjugate regions to the upper and lower right of $q_{inner}$,
$\lambda_{Z,L,5}$ are dominant.  In general, as the temperature $T$ increases 
from 0 to infinity for the
Potts antiferromagnet, i.e. as $a$ increases from 0 to 1, the locus ${\cal B}$
moves in toward the origin in the $q$ plane and degenerates to a point at the
origin for $a=1$.

\begin{figure}[hbtp]
\centering
\leavevmode
\epsfxsize=2.5in
\begin{center}
\leavevmode
\epsffile{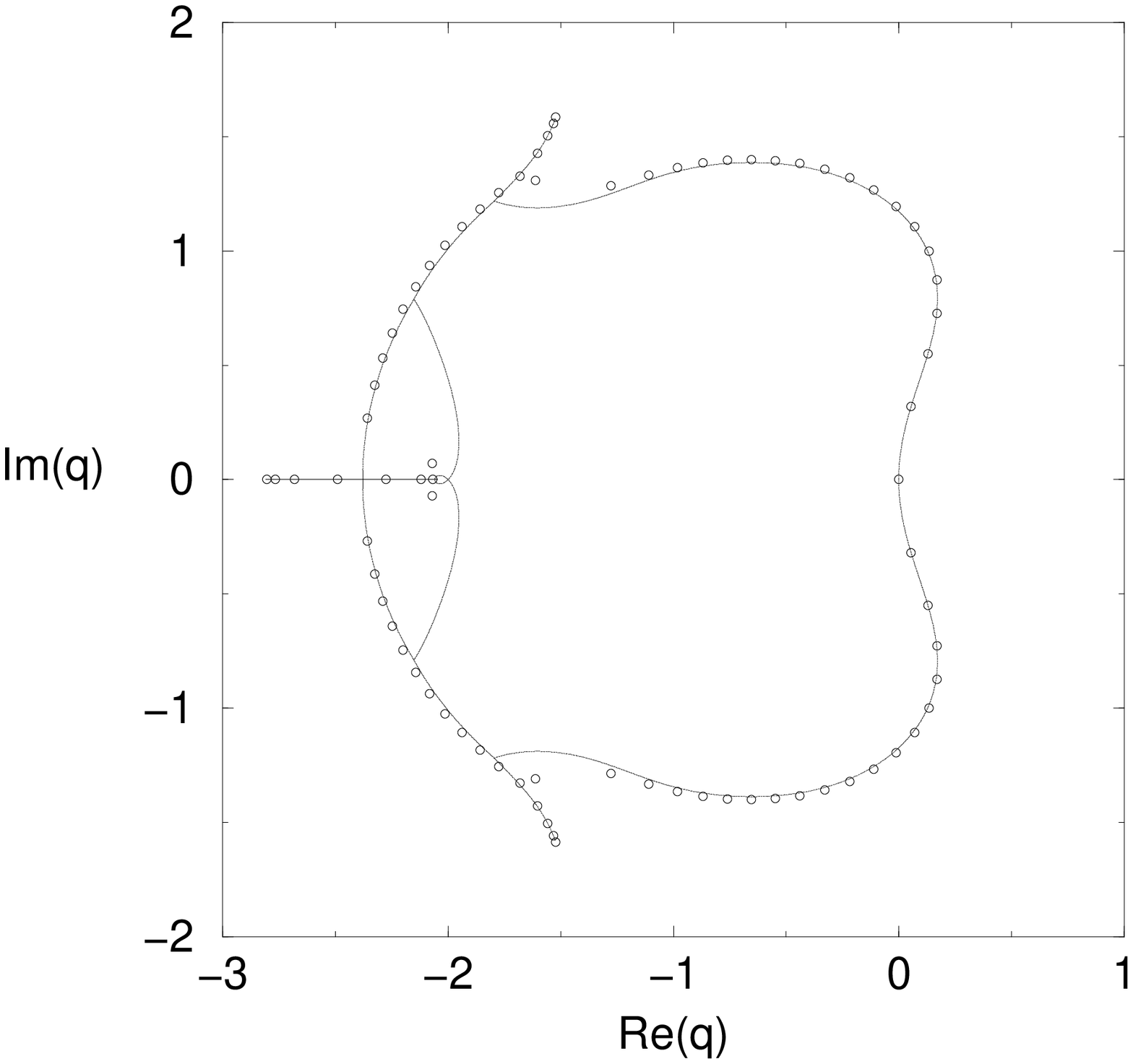}
\end{center}
\caption{\footnotesize{Locus ${\cal B}_q(\{L\})$ for the Potts ferromagnet on
the $m \to \infty$ limit of the cyclic or M\"obius $L_y=2$ strip of the
honeycomb lattice for $a=2$. For comparison, chromatic zeros are shown for
the cyclic strip with $m=20$ (i.e., $n=80$).}}
\label{hpxy2a2}
\end{figure}

The boundary ${\cal B}_q$ for the Potts ferromagnet is given for the
illustrative finite-temperature value $a=2$ in Fig. \ref{hpxy2a2}.  Compared
with analogous figures for the infinite-length limits of the $L_y=2$ strips of
the square and triangular lattices (Figs. 13 in \cite{a} and \cite{ta}), one
sees the common feature that ${\cal B}_q$ passes through $q=0$, does not cross
the positive real $q$, but does have complex-conjugate lobes that extend into
the $Re(q) > 0$ half-plane.  Similar to ${\cal B}_q$ for this strip of the
square lattice, here ${\cal B}_q$ contains a finite line-segment on the
negative $q$ axis, whereas such a line segment is not present on ${\cal B}_q$
for the triangular strip for this value of $a$.

\subsection{${\cal B}$ in the Complex-Temperature Plane} 

\begin{figure}[hbtp]
\centering
\leavevmode
\epsfxsize=2.5in
\begin{center}
\leavevmode
\epsffile{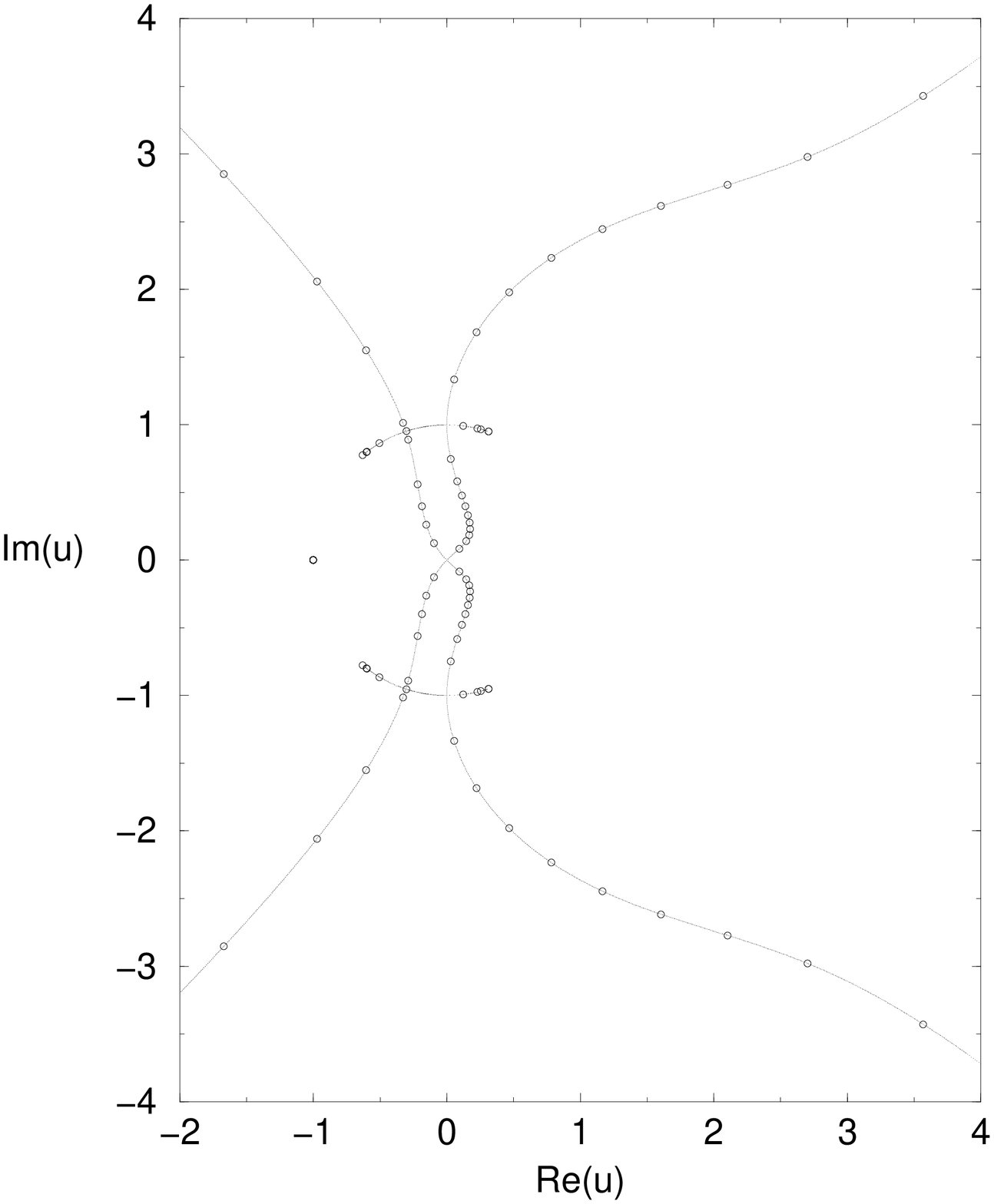}
\end{center}
\caption{\footnotesize{Locus ${\cal B}_u(\{L\})$ for the $m \to \infty$ limit
of the cyclic strip of the honeycomb lattice with $q=2$. Partition function
zeros are shown for $m=20$, so that $Z$ is a polynomial of degree
$e=5m=100$ in $v$ and hence, up to an overall factor, in $u$).}}
\label{hpxy2q2}
\end{figure}

\begin{figure}[hbtp]
\centering
\leavevmode
\epsfxsize=2.5in
\begin{center}
\leavevmode
\epsffile{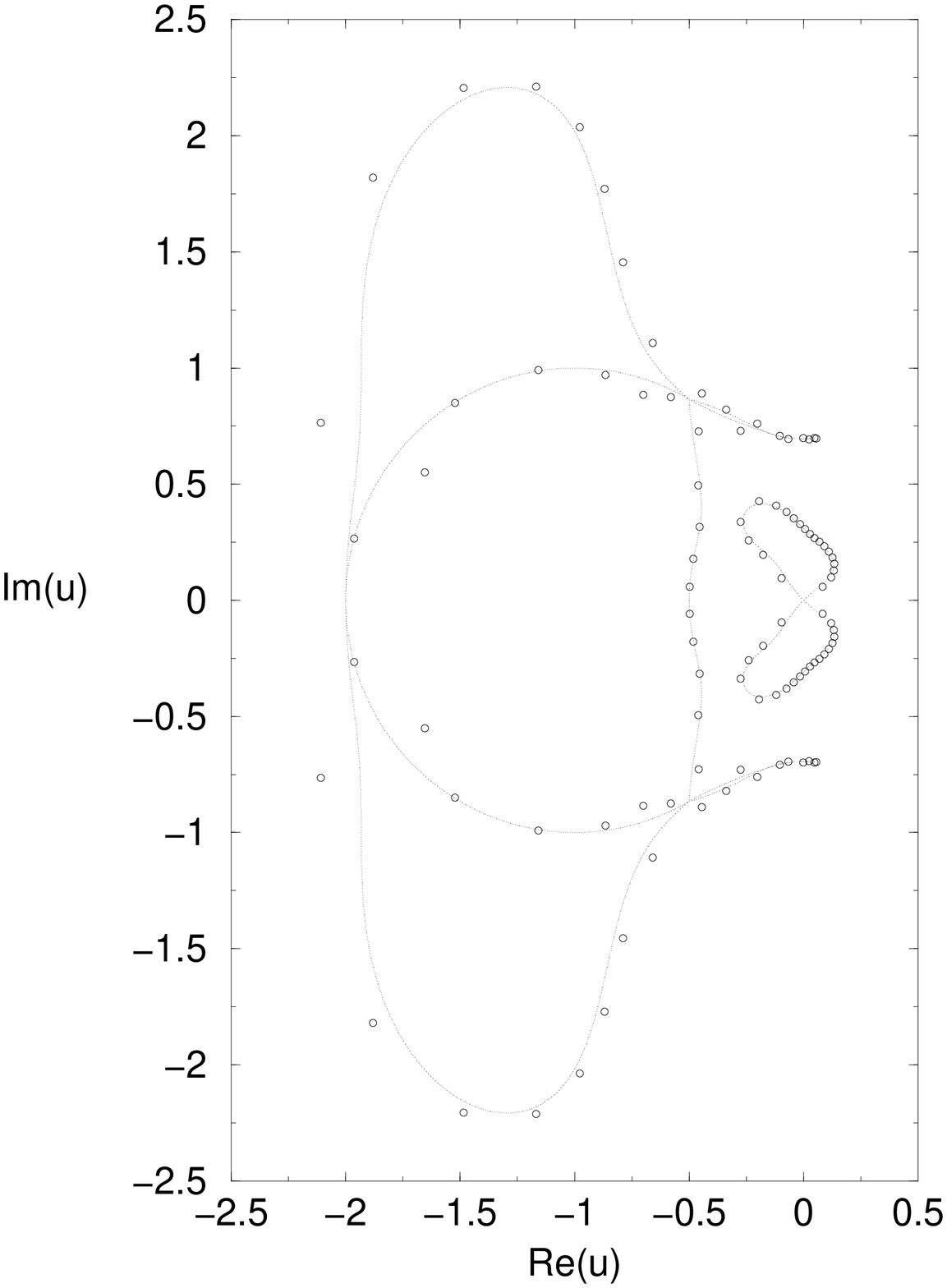}
\end{center}
\caption{\footnotesize{Locus ${\cal B}_u(\{L\})$ for $q=3$. Partition function
zeros are shown for $e=100$, as in Fig. \ref{hpxy2q2}.}}
\label{hpxy2q3}
\end{figure}

\begin{figure}[hbtp]
\centering
\leavevmode
\epsfxsize=2.5in
\begin{center}
\leavevmode
\epsffile{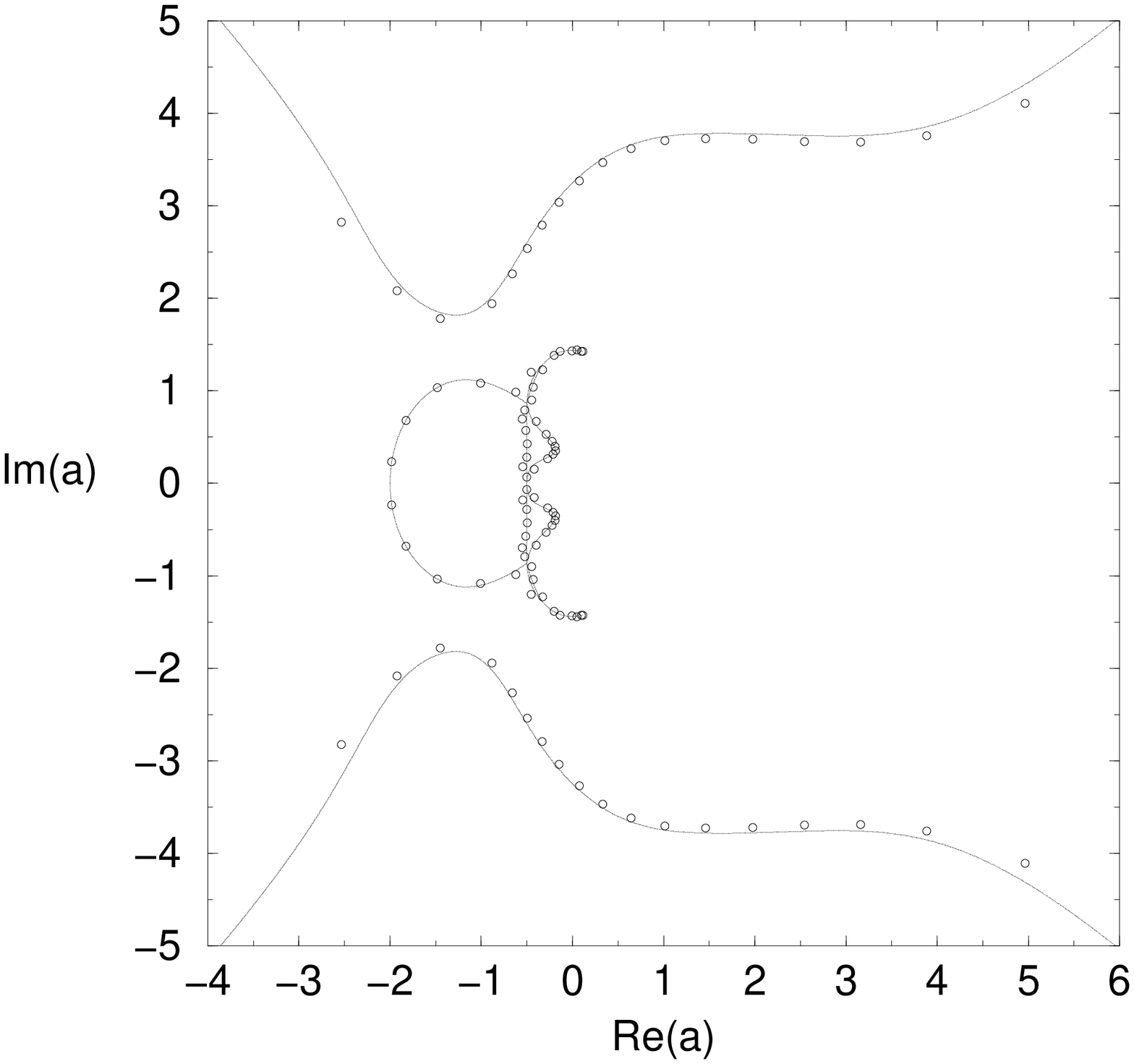}
\end{center}
\caption{\footnotesize{Locus ${\cal B}_a(\{L\})$ for $q=3$. Partition function 
zeros are shown for $e=100$.}}
\label{hpxy2aq3}
\end{figure}

\begin{figure}[hbtp]
\centering
\leavevmode
\epsfxsize=2.5in
\begin{center}
\leavevmode
\epsffile{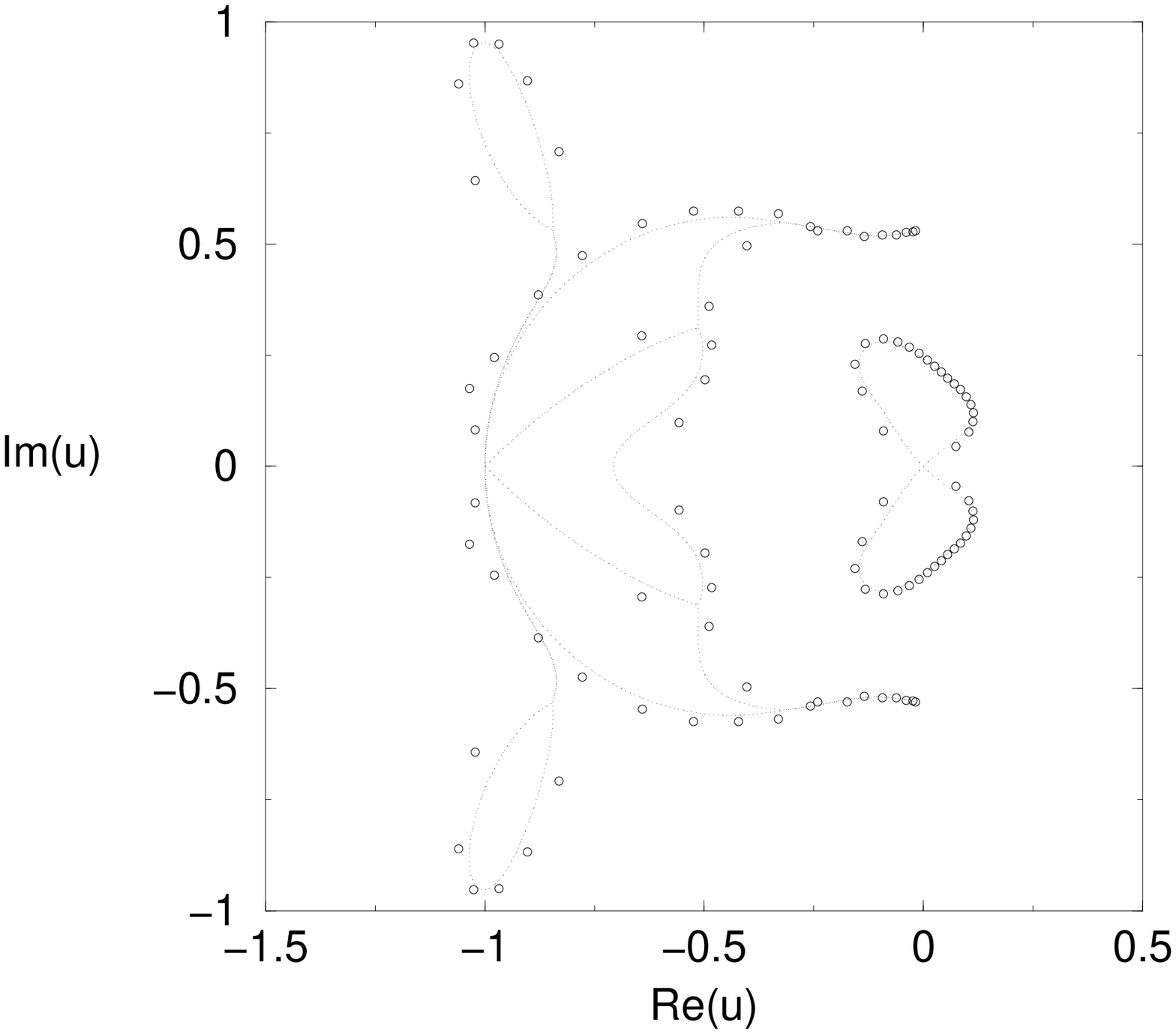}
\end{center}
\caption{\footnotesize{Locus ${\cal B}_u(\{L\})$ for $q=4$. Partition function
zeros are shown for $e=100$.}}
\label{hpxy2q4}
\end{figure}

\begin{figure}[hbtp]
\centering
\leavevmode
\epsfxsize=2.5in
\begin{center}
\leavevmode
\epsffile{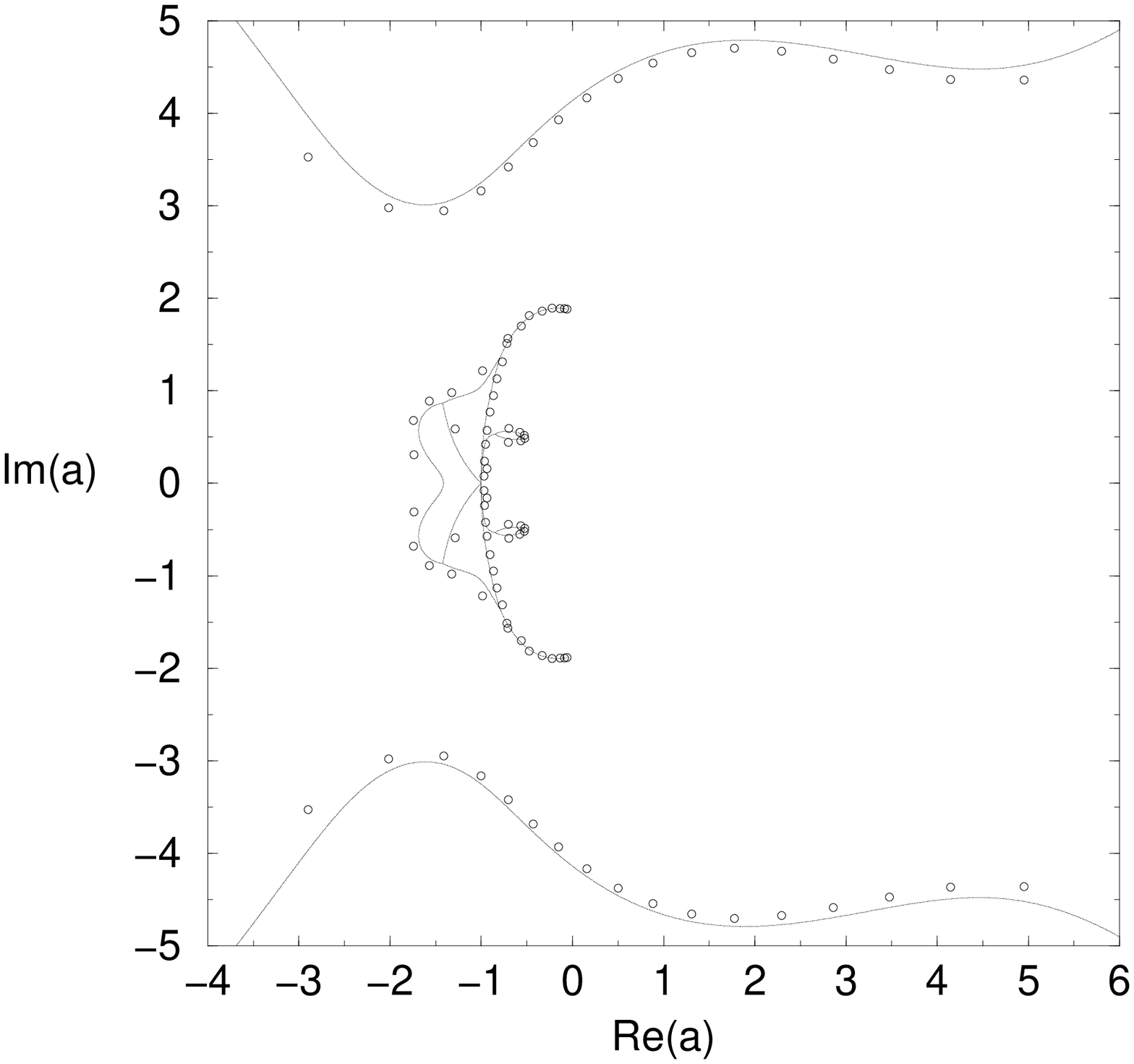}
\end{center}
\caption{\footnotesize{Locus ${\cal B}_a(\{L\})$ for $q=4$. Partition function
zeros are shown for $e=100$.}}
\label{hpxy2aq4}
\end{figure}

\begin{figure}[hbtp]
\centering
\leavevmode
\epsfxsize=2.5in
\begin{center}
\leavevmode
\epsffile{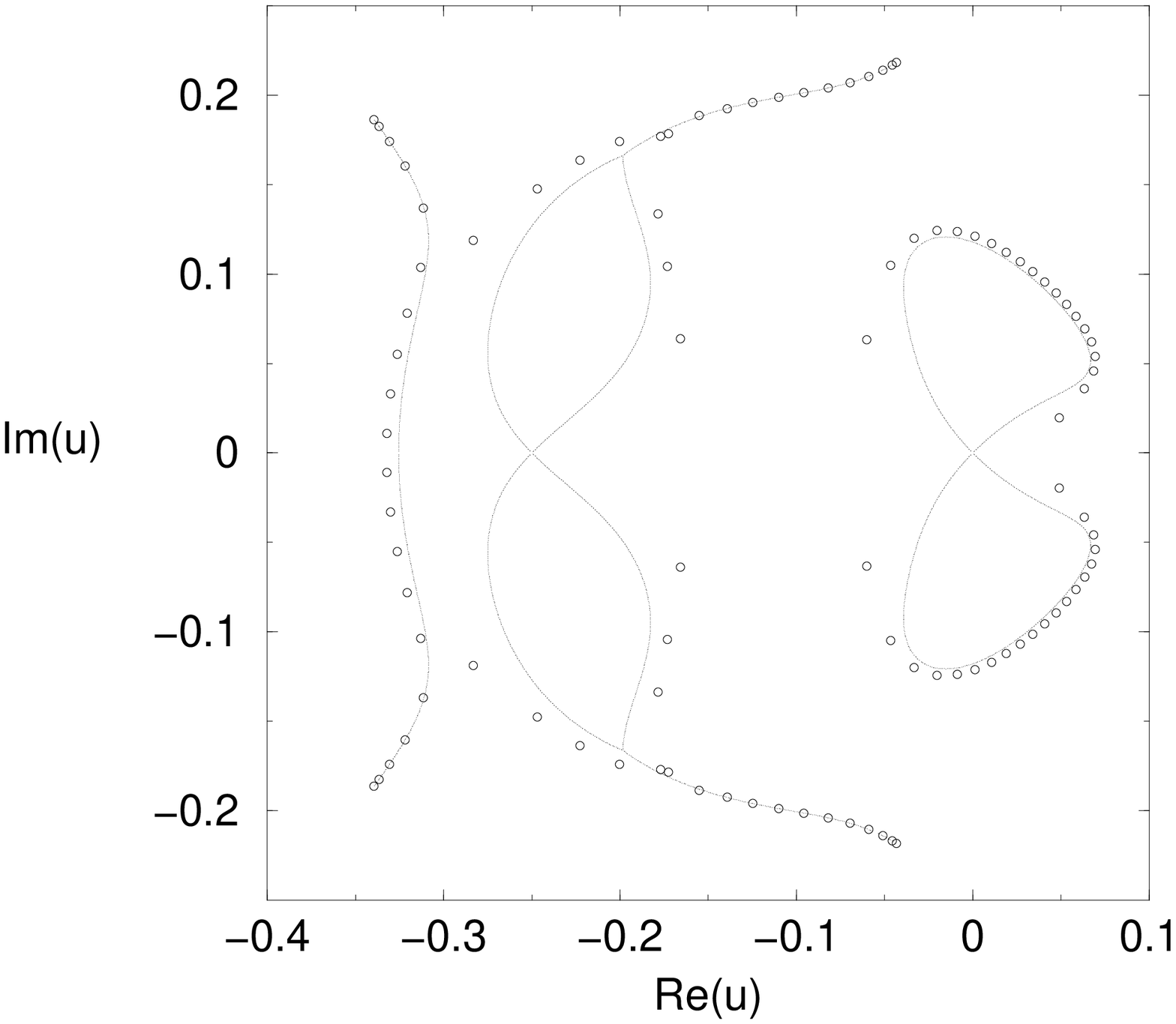}
\end{center}
\caption{\footnotesize{Locus ${\cal B}_u(\{L\})$ for $q=10$. Partition function
zeros are shown for $e=100$.}}
\label{hpxy2q10}
\end{figure}

We next discuss slices of ${\cal B}$ in the complex-temperature plane, as a
function of $q$. In Figs. \ref{hpxy2q2}-\ref{hpxy2q10} we show plots of
${\cal B}$ for $q=2$, 3, 4, and 10.  The locus ${\cal B}$ is noncompact in the
$a$ plane and, for $q \ne 2$, compact in the $u$ plane. For $q=2$ there can be
noncommutativity of the type (\ref{fnoncomm}); the results that we show are for
the $f_{nq}$ definition, i.e., we set $q=2$ first before taking the limit $n
\to \infty$.  For comparison with previous calculations of complex-temperature
zeros in \cite{p,p2}, we show some plots in the $a$ plane as well as $u$
plane. 

In the $u$ plane, four curves forming two branches of ${\cal B}$ cross at
$u=0$, reflecting the $T=0$ critical point of the $q$-state Potts ferromagnet.
In general, the density of complex-temperature (i.e., Fisher \cite{fisher}) 
zeros along the curves comprising ${\cal B}$ in the
vicinity of a generic singular point $u_s$ behaves as \cite{abe}
\beq
g \sim |u-u_s|^{1-\alpha_s}
\label{gdensity}
\eeq
where $\alpha_s$ ($\alpha_s'$) denotes the
corresponding specific heat exponent for the approach to $u_s$ from within the
CTE PM (FM) phase.  Thus, for a continuous, second-order transition, with
$\alpha_s < 1$, this density vanishes as one approaches the critical point
$u_s$ along ${\cal B}$, while, in contrast, if $\alpha_s=1$, corresponding to a
first-order transition, this density would remain nonzero as the boundary
crosses $u_s$, and there would be a discontinuity in the internal energy, $U$.
(We recall that this is analogous to the discontinuity in $M(H)$ for $T < T_c$
as one changes $H$ from positive through zero to negative values, and the
connection to the nonzero density of Yang-Lee (complex-field) zeros, via the
relation $M(T)=2\pi g(H=0)$, where $M(T)$ is the spontaneous magnetization
\cite{yl,ly}.)  Now the essential zero in the specific heat in (\ref{cfmlow})
at $T=0$, if expressed in terms of an algebraic specific exponent $\alpha$, 
corresponds to $\alpha=-\infty$ at $u=u_s=0$.  Substituting this into
(\ref{gdensity}), it follows that the density vanishes rapidly as one 
approaches the origin $u=0$ along the curves forming ${\cal B}_u$.  This is 
evident in Figs. \ref{hpxy2q2} and \ref{hpxy2q3}. 
The angles at which these curves cross the origin are $\pm \pi/4$ and $\pm
3\pi/4$, as is clear in both of the figures shown, for $q=2$ and 3.  In
particular, in the Ising case $q=2$, because of the bipartite property of the
cyclic strip of the honeycomb lattice, the antiferromagnetic and ferromagnetic
versions of the model are equivalent, and so the antiferromagnet also has a
$T=0$ critical point.  This is manifested in the fact that ${\cal B}_a$ passes
through the origin of the $a$ plane and ${\cal B}_q$ passes through the point
$q=2$. Indeed, for this value $q=2$, the locus ${\cal B}$ is the same in the
$a$ and $u$ plane, i.e. it is invariant under the inversion map $a \to
a^{-1}=u$.  As was the case with the phase diagram for the infinite-length
limit of the open strip, the positive $u$ axis and its maximal
complex-temperature analytic continuation forms the (CTE)PM phase.  As
discussed earlier \cite{bcc,a,s4}, the fact that, for $a=0$, 
the crossing of ${\cal B}_q$
at $q=2$ signals the property that the Ising antiferromagnet on the
finite-width, infinite-length strips of the square lattice has a $T=0$ critical
point is a useful feature of the periodic or reversed-orientation periodic
longitudinal boundary conditions; in contrast, the loci ${\cal B}_q$ at $a=0$
for strips with free longitudinal boundary conditions do not, in general, pass
through $q=2$.  Furthermore, just as we showed by exact calculations in the
case with the strips of the square lattice, in the present case of the
honeycomb lattice, we anticipate that at the antiferromagnetic zero-temperature
value $a=0$, i.e., $v=-1$, the loci ${\cal B}_q$ for wider cyclic/M\"obius 
strips will exhibit at least 
three crossings: (i) at a value $q_c$ that approaches the value for the
infinite honeycomb lattice 
\beq
q_c(hc)=\frac{3+\sqrt{5}}{2}=2.61803...
\label{qchc}
\eeq
(the relevant root of eq. (\ref{veq}) for $v=-1$, as discussed below), 
(ii) at $q=2$, reflecting the $T=0$ critical
point of the Ising antiferromagnet on these strips, and (iii) at $q=0$.

One of the reasons for interest in the complex-temperature phase diagrams for
the $q$-state Potts model on infinite-length, finite-width strips of various
lattices is that these provide exact results that exhibit features which can
give insight into the properties of the complex-temperature phase diagrams of
the corresponding Potts model on the respective 2D lattices \cite{a,ta}.  Of
course for the $q=2$ Ising case, since the model is solved, this
complex-temperature phase diagram is known \cite{chihc}.  However, for other
values of $q$, these insights can help one in interpreting the information that
one has from calculations of complex-temperature zeros on finite sections of
the honeycomb lattice \cite{p,p2}.  In order to appreciate some of the similar
features in the complex-temperature phase diagrams, we recall that ${\cal B}_a$
for the Ising model on the honeycomb lattice consists of an arc of the unit
circle $a=e^{\pm i\theta}$ with $\pi/3 \le \theta \le \pi$, together with a
closed, bean-shaped curve that intersects the real axis at $a=2-\sqrt{3}$, the
paramagnetic-antiferromagnetic critical point, and its inverse, $a=2+\sqrt{3}$,
the paramagnetic-ferromagnetic critical point \cite{chihc}.  The arc of the
unit circle and the closed curve intersect at $a=\pm i$.  There are, of course,
obvious qualitative differences between the locus ${\cal B}_a$ for the
finite-width, infinite-length strip and the full 2D honeycomb lattice,
reflecting the fact that the 2D Ising ferromagnetic (and equivalently, the
antiferromagnet) has a finite-temperature critical point, while this occurs
only at zero-temperature on the infinite-length, finite-width strip, which is
effectively a quasi-one-dimensional system.  However, for both the strip and
the full honeycomb lattice, ${\cal B}_a$ includes intersection points where two
branches of curves cross at $a=\pm i$.  The locus ${\cal B}_a$ for the
infinite-length $L_y=2$ strip also exhibits complex-conjugate arc endpoints in
the $Re(a) > 0$ half-plane, as is true of the exactly solved Ising model on the
full 2D honeycomb lattice; in addition, this locus for the strip exhibits
complex-conjugate arc endpoints in the $Re(a) < 0$ half-plane.  This
illustrates how, although physical temperature properties of the Potts model on
an finite-width, infinite strip of a given lattice and on the full 2D lattice
are clearly different, complex-temperature properties can have some
similarities.

Having discussed the situation with the $q=2$ case, where the 2D model is
exactly solved, let us proceed to cases where the 2D Potts model has not been
solved.  In general, for the honeycomb lattice, from duality and a 
star-triangle relation, one can derive an algebraic equation for the PM-FM 
and PM-AFM critical points \cite{wurev,hccrit1,hccrit2}
\beq
v^3-3qv-q^2=0 \ .
\label{veq}
\eeq 
(where $v=a-1$). 
For example, for $q=3$, the equation yields three solutions (tabulated, e.g.,
as eqs. (3.3)-(3.5) in \cite{p}), namely (i) the PM-FM critical point at 
$a_{PM-FM,q=3}=1+2\sqrt{3}\cos(\pi/18)=4.41147...$, and two complex-temperature
singular points, at $a=1-\sqrt{3}\cos(\pi/18)+3\sin(\pi/18)=-0.1847925...$ 
and $a=1-\sqrt{3}\cos(\pi/18)-3\sin(\pi/18)=-1.22668...$.  There is no physical
PM-AFM critical point solution, and the absence of such a phase transition was
confirmed by explicit Monte Carlo study of the model in \cite{pafhc3} (and
further confirmed in \cite{salas}). More generally, considering this solution 
as a 
function of $q$ generalized from positive integer to real values, one sees that
it decreases from the $q=2$ Ising value $a=\sqrt{2}-1$ and passes through $a=0$
at the value $q_c(hc)$ given in eq. (\ref{qchc}) above (see the middle curve in
Fig. 4 of \cite{p}), corresponding to a zero-temperature critical point for the
Potts antiferromagnet at this value of $q$.  For $q_c(hc) \le q \le 4$, this 
root decreases from 0 to $-1$, where it coalesces with another solution (the
lower curve in Fig. 4 of \cite{p}) to form a double root, and for $q > 4$,
eq. (\ref{veq}) has only a single real root, namely the PM-FM critical point. 
In \cite{hcl}, as a rigorous result of duality, it was shown that as a
consequence of the (first-order) phase transition of the $q=3$ Potts 
antiferromagnet on the triangular lattice at $a_{PM-AFM,t} \simeq 0.203$, 
it follows that for $q=3$ Potts model on the honeycomb lattice, the left-most 
crossing of the locus ${\cal B}_a$ is at the complex-temperature point given by
the duality mapping $v \to q/v$, i.e., 
\beq
a \to \frac{q+a-1}{a-1}
\label{adual}
\eeq
i.e., $a \simeq -2.76$. 
The calculations of complex-temperature zeros in \cite{p} (given in Figs. 5-7
of that paper) were consistent with these 
crossings of ${\cal B}_a$ at $a \simeq -2.76$, $-1.23$, and $-0.185$ and also 
suggesteda fourth possible crossing, at $a \simeq -0.65$, for a total of four 
crossings on the negative real $a$ axis.  

For the $q=3$ Potts model on the infinite-length strip, as shown in
Fig. \ref{hpxy2aq3}, we find that ${\cal B}_a$ crosses the negative real $a$
axis at two points, $a=-1/2$ and $a=-2$.  In order to relate this
complex-temperature phase diagram to one for the $q=3$ Potts model on the 2D
honeycomb lattice, one can imagine performing a transformation in which one
retracts the branches of the curves forming ${\cal B}_a$ from the
zero-temperature ferromagnet critical point at $u=0$; in the right-hand
half-plane, one would then reconnect the two branches of the curve so as to
cross the positive real $a$ axis at the value of $a_{PM-FM,q=3}$ given above,
while in the left-hand half plane, one would reconnect the branches to cross at
a point on the negative real $a$ axis.  One could also pull apart the two
curves that have a tacnode multiple point at $a=-1/2$ to obtain a total of four
crossings on the negative real axis, as appears to be the case for the model on
the full 2D honeycomb lattice.  The exactly determined locus ${\cal B}_a$ also
exhibits a complex-conjugate pair of arc endpoints in the $Re(a) > 0$
half-plane, qualitatively similar to the arc endpoints that appear to be formed
by the complex-temperature zeros for 2D honeycomb lattice.  Thus, although the
physical thermodynamic properties are clearly different, certain
complex-temperature features of the phase diagram for the $q=3$ Potts model on
the infinite-length, width $L_y=2$ strip of the honeycomb lattice show
interesting qualitative similarities to those inferred for the
complex-temperature phase diagram of the model on the full honeycomb lattice.

We next consider the case $q=4$.  For this value, eq. (\ref{veq}) reduces, in
terms of the variable $a$, to $(a-5)(a+1)^2=0$, yielding the physical PM-FM
critical point $a_{PM-FM,q=4}=5$ and a double root at the complex-temperature
point $a=u=-1$.  As in the $q=3$ case, there is no phase transition at finite
or zero temperature for the $q=4$ Potts antiferromagnet. From the duality
relation given in \cite{hcl} together with the fact that the $q=4$ Potts
antiferromagnet has a $T=0$ critical point on the triangular lattice, it
follows that the left-most crossing of ${\cal B}_a$ on the real axis for the
$q=4$ Potts model on the honeycomb lattice is at the value of $a$ given by
(\ref{adual}), viz., $a=-3$.  The calculations of complex-temperature zeros in
\cite{p2} (see Figs. 2-4 of that paper) exhibited these three crossings of the
complex-temperature phase boundary ${\cal B}_a$ at $a=5$, $a=-1$ and $a=-3$ and
also suggested crossings at approximately $a=-2$ and $a=-1/2$.  The locus
${\cal B}_a$ also exhibited a pair of arc endpoints (prongs) in the 
$Re(a) > 0$ half-plane and structures that resembled a complex-conjugate pair
of prongs or bulbs in the $Re(a) < 0$ half-plane, all within the
interior of the PM phase.

What insights can we draw for this $q=4$ case from our exact results for the
infinite-length, $L_y=2$ strip?  The locus ${\cal B}$ is shown in the $u$ and
$a$ planes in Figs. \ref{hpxy2q4} and \ref{hpxy2aq4}.  In the $a$ plane, we
again perform a transformation to remove the intrinsically 1D features of the
complex-temperature phase boundary, i.e. their passage through $u=0$ for the
comparison with the 2D results. We thus retract the branches of the curves
passing through this point; in the right-hand half-plane, we reconnect the two
branches of the curve so as to cross the positive real $a$ axis at the value of
$a_{PM-FM,q=4}$ given above, while in the left-hand half plane, we reconnect
the branches to cross at a point on the negative real $a$ axis, which then
forms the left-most crossing.  We could also pull apart the curves that meet
and cross at $a=-1$, thereby obtaining four crossings of ${\cal B}_a$ on the
negative real $a$ axis.  The two complex-conjugate pairs of arc endpoints on
${\cal B}_a$ are in nice 1-1 correspondence with those observed for the model
on the 2D honeycomb lattice in \cite{p2}.  In addition, ${\cal B}_a$ exhibits
two small bulb-like structures around $a=-0.65 \pm 0.53i$; these bear an
interesting correspondence to the analogous structures found in \cite{p2}.
This discussion shows how some qualitative aspects of the complex-temperature
phase diagram for the Potts model on the 2D honeycomb lattice can correspond to
those found via exact solutions for infinite-length, finite-width strips.  One
of the most remarkable findings is that the crossing point of ${\cal B}_a$ at
$a=u=-1$ for the infinite-length strip is not just qualitatively but exactly
the same as one finds, rigorously, from the solution of (\ref{veq}), for ${\cal
B}_a$ on the infinite honeycomb lattice.  This is also true for the infinite
triangular lattice \cite{p2}, since it is dual to the honeycomb lattice, and
under the duality mapping (\ref{adual}), $a=-1$ maps to itself if $q=4$.  As
was discussed in \cite{pfef}, the point $a=-1$ is a solution of the equation
for the critical manifold, $v^2=q$, at $q=4$ and hence ${\cal B}_a$ passes
through $a=-1$ for the square lattice.  Calculations of complex-temperature
zeros for the $q=4$ Potts model on large finite sections of the triangular
lattice in \cite{p2} and the square lattice (see \cite{wuetal,wuz,pfef} and
references therein) are both consistent with these exact results.  Moreover,
our exact determinations of ${\cal B}_a$ for the 1D Potts model (i.e., $n \to
\infty$ limit of the circuit graph) \cite{is1d,a} and infinite-length, $L_y=2$
cyclic/M\"obius and open strips of the square and triangular lattices
\cite{a,ta} showed that in all cases, if $q=4$, then ${\cal B}_a$ passes
through $a=-1$.  It is of interest to see what the values $q=4$, $a=v+1=-1$
correspond to in terms of the variables $x$ and $y$, defined in
eqs. (\ref{xdef}) and (\ref{ydef}), that enter in the Tutte polynomial
equivalent to the Potts model partition function (see eqs.  (\ref{tuttepol})
and (\ref{ztutte})).  We find that $(q,v)=(4,-2)$ corresponds to the symmetric
point $(x,y)=(-1,-1)$.  More generally, one can investigate the continuous
accumulation set of zeros of the Tutte polynomial in the complex $x$ plane for
fixed $y$, ${\cal B}_x$ and in the complex $y$ plane for fixed $x$, ${\cal
B}_y$. Since this information is equivalent to our study of the loci ${\cal
B}_q$ for fixed $u$ and ${\cal B}_u$ for fixed $q$, we shall not include the
results here.

An illustration of ${\cal B}_u$ for larger values of $q$ is given for the case
$q=10$ in Fig. \ref{hpxy2q10}.  This locus includes two separate figure-eight
type curves, the one to the left also involving outgoing arcs, and, to the far
left, a self-conjugate arc.  As compared with our result for the
infinite-length $L_y=2$ strip of the square lattice in Fig. 17 of \cite{a}, one
sees that the middle figure-eight curve plus outlying arcs that are present on
${\cal B}_u$ for the honeycomb strip are absent in the locus for the strip of
the square lattice, while the other two components (the self-conjugate arc on
the left and the figure-eight curve on the right) are qualitatively similar.
For the infinite-length honeycomb lattice strip with open boundary conditions,
we note that ${\cal B}_u$ consists of the same self-conjugate arc on the left
as in Fig. \ref{hpxy2q10}, together with two complex-conjugate pairs of arcs.

\section{$L_y=3$ Cyclic and M\"obius Strips of the Honeycomb Lattice} 

\begin{figure}[hbtp]
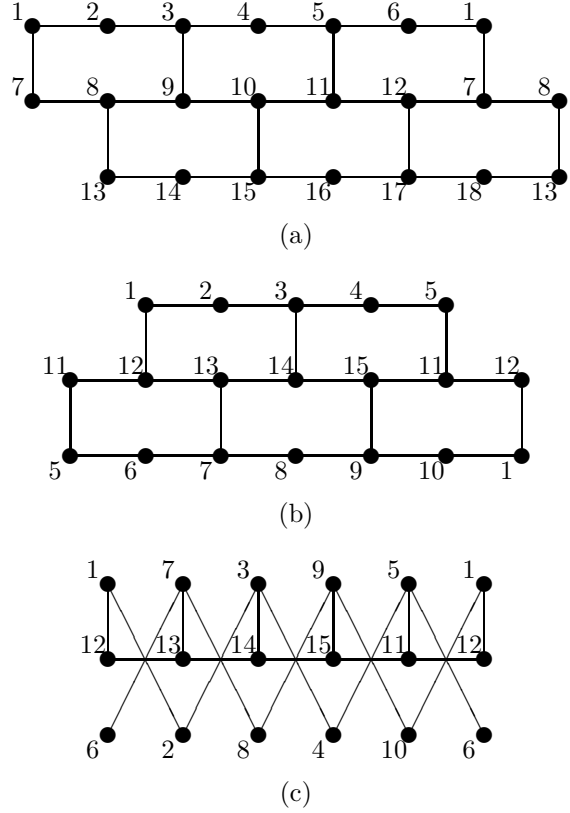

\caption{\footnotesize{Illustrative strip graphs of the honeycomb lattice,
displayed as a brick lattice, with width $L_y=3$: (a) cyclic,
length $m=3$ and (b) M\"obius, length $m=2$.  We also show
(c) an $L_y=3$ cyclic crossing-subgraph strip of the honeycomb lattice with
length $\ell=5$.  As discussed in the text, the cyclic crossing-subgraph strips
of a given width are equivalent to the cyclic and M\"obius strips of the same
width for even and odd $\ell$, respectively.  This equivalence is illustrated
here; graph (c) is equivalent to (b).}}
\label{hcstrips}
\end{figure}

Illustrations of strips of the honeycomb lattice with width $L_y=3$ and 
$(FBC_y,PBC_x)=$ cyclic and $(FBC_y,TPBC_x)=$ M\"obius boundary conditions,
displayed as strips of a brick lattice, are shown in Fig. \ref{hcstrips}
(a,b).  

\vspace*{1cm}
\unitlength 1mm
\begin{center}
\begin{picture}(70,20)
\multiput(10,0)(10,0){7}{\circle*{2}}
\multiput(0,10)(10,0){8}{\circle*{2}}
\multiput(0,20)(10,0){7}{\circle*{2}}
\multiput(10,0)(20,0){4}{\line(0,1){10}}
\multiput(0,10)(20,0){4}{\line(0,1){10}}
\put(10,0){\line(1,0){60}}
\put(0,10){\line(1,0){70}}
\put(0,20){\line(1,0){60}}
\put(8,-2){\makebox(0,0){13}}
\put(18,-2){\makebox(0,0){14}}
\put(28,-2){\makebox(0,0){15}}
\put(38,-2){\makebox(0,0){16}}
\put(48,-2){\makebox(0,0){17}}
\put(58,-2){\makebox(0,0){18}}
\put(68,-2){\makebox(0,0){13}}
\put(-2,12){\makebox(0,0){7}}
\put(8,12){\makebox(0,0){8}}
\put(18,12){\makebox(0,0){9}}
\put(28,12){\makebox(0,0){10}}
\put(38,12){\makebox(0,0){11}}
\put(48,12){\makebox(0,0){12}}
\put(58,12){\makebox(0,0){7}}
\put(68,12){\makebox(0,0){8}}
\put(-2,22){\makebox(0,0){1}}
\put(8,22){\makebox(0,0){2}}
\put(18,22){\makebox(0,0){3}}
\put(28,22){\makebox(0,0){4}}
\put(38,22){\makebox(0,0){5}}
\put(48,22){\makebox(0,0){6}}
\put(58,22){\makebox(0,0){1}}
\put(35,-8){\makebox(0,0){(a)}}
\end{picture}
\end{center}
\vspace*{1cm}

\begin{center}
\begin{picture}(60,20)      
\multiput(0,0)(10,0){7}{\circle*{2}}
\multiput(0,10)(10,0){7}{\circle*{2}}
\multiput(10,20)(10,0){5}{\circle*{2}}
\multiput(0,0)(20,0){4}{\line(0,1){10}}
\multiput(10,10)(20,0){3}{\line(0,1){10}}
\put(10,20){\line(1,0){40}}
\multiput(0,0)(0,10){2}{\line(1,0){60}}
\put(-2,-2){\makebox(0,0){5}}
\put(8,-2){\makebox(0,0){6}}
\put(18,-2){\makebox(0,0){7}}
\put(28,-2){\makebox(0,0){8}}
\put(38,-2){\makebox(0,0){9}}
\put(48,-2){\makebox(0,0){10}}
\put(58,-2){\makebox(0,0){1}}
\put(-2,12){\makebox(0,0){11}}
\put(8,12){\makebox(0,0){12}}
\put(18,12){\makebox(0,0){13}}
\put(28,12){\makebox(0,0){14}}
\put(38,12){\makebox(0,0){15}}
\put(48,12){\makebox(0,0){11}}
\put(58,12){\makebox(0,0){12}}
\put(8,22){\makebox(0,0){1}}
\put(18,22){\makebox(0,0){2}}
\put(28,22){\makebox(0,0){3}}
\put(38,22){\makebox(0,0){4}}
\put(48,22){\makebox(0,0){5}}
\put(30,-8){\makebox(0,0){(b)}}
\end{picture}
\end{center}
\vspace*{1cm}

\begin{center}
\begin{picture}(50,20)
\multiput(0,0)(10,0){6}{\circle*{2}}
\multiput(0,10)(10,0){6}{\circle*{2}}
\multiput(0,20)(10,0){6}{\circle*{2}}
\multiput(0,0)(10,0){5}{\line(1,2){10}}
\multiput(0,20)(10,0){5}{\line(1,-2){10}}
\put(0,10){\line(1,0){50}}
\multiput(0,10)(10,0){6}{\line(0,1){10}}
\put(-2,-2){\makebox(0,0){6}}
\put(8,-2){\makebox(0,0){2}}
\put(18,-2){\makebox(0,0){8}}
\put(28,-2){\makebox(0,0){4}}
\put(38,-2){\makebox(0,0){10}}
\put(48,-2){\makebox(0,0){6}}
\put(-2,12){\makebox(0,0){12}}
\put(8,12){\makebox(0,0){13}}
\put(18,12){\makebox(0,0){14}}
\put(28,12){\makebox(0,0){15}}
\put(38,12){\makebox(0,0){11}}
\put(48,12){\makebox(0,0){12}}
\put(-2,22){\makebox(0,0){1}}
\put(8,22){\makebox(0,0){7}}
\put(18,22){\makebox(0,0){3}}
\put(28,22){\makebox(0,0){9}}
\put(38,22){\makebox(0,0){5}}
\put(48,22){\makebox(0,0){1}}
\put(25,-8){\makebox(0,0){(c)}}
\end{picture}
\end{center}

We calculate the chromatic polynomials by iterated use of the
deletion-contraction theorem and by coloring matrix methods. Here and below,
in our results for chromatic polynomials, we shall suppress the subscript $P$
in $\lambda_{P,G,j}$ and $c_{P,G,j}$ where it is obvious. 
For the $L_y=3$ cyclic and M\"obius strips of the honeycomb lattice, denoted 
$hc3c=hc,3 \times m, FBC_y,PBC_x$ and $hc3mb=hc,3 \times m, FBC_y,TPBC_x$, we 
obtain the result $N_{hc3c,\lambda}=N_{hc3mb,\lambda}=14$ and 
\beq
P(hc,3 \times m,FBC_y,PBC_x,q) = \sum_{j=1}^{14} c_{hc3c,j} 
(\lambda_{hc3,j})^m
\label{phc3cyc}
\eeq
\beq
P(hc,3 \times m,FBC_y,TPBC_x,q) = \sum_{j=1}^{14} c_{hc3mb,j}
(\lambda_{hc3,j})^m
\label{phc3mob}
\eeq
where, for $j=1,2$, 
\beq
\lambda_{hc3,1}=1
\label{lamhc1}
\eeq
and
\beq
\lambda_{hc3,2}=(q-1)^2 \ . 
\label{lamhc2}
\eeq
Note that 
\beq
\lambda_{hc3,j}=\lambda_{hc2,j} \quad {\rm for} \ \ j=1,2 \ . 
\label{lameq12}
\eeq
The $\lambda_{hc3,j}$ for $j=3,4,5$ are the roots of the cubic equation
\beq
\xi^3-(2q^2-8q+11)\xi^2+(q^4-8q^3+26q^2-38q+23)\xi-(q-1)^2=0 
\label{eqcub1}
\eeq
and 
the $\lambda_{hc3,j}$ for $6 \le j \le 11$ are the roots of the sixth degree
equation
\beqs
& & \xi^6-(3q^4-18q^3+46q^2-60q+37)\xi^5 \cr\cr
& & +(3q^8-36q^7+199q^6-662q^5+1456q^4-2172q^3+2144q^2-1278q+358)\xi^4 \cr\cr
& & -(q^{12}-18q^{11}+154q^{10}-828q^9+3117q^8-8646q^7+18070q^6-28564q^5+
33782q^4 \cr\cr
& & -29110q^3+17370q^2-6466q+1142)\xi^3 \cr\cr
& & (q-1)^2(q^{12}-20q^{11}+186q^{10}-1060q^9+4115q^8-11454q^7+23448q^6 \cr\cr
& & -35642q^5+40094q^4-32780q^3+18680q^2-6756q+1197)\xi^2 \cr\cr
& & -(q-1)^4(2q^8-26q^7+150q^6-498q^5+1037q^4-1388q^3+1176q^2-590q+141)\xi 
\cr\cr 
& & +(q-2)^2(q-1)^8=0 \ . 
\label{eqsix}
\eeqs
Finally, $\lambda_{hc3,j}$ for $12 \le j \le 14$ are roots of the cubic
equation 
\beqs
& & \xi^3-(q^6-8q^5+28q^4-56q^3+71q^2-58q+26)\xi^2 \cr\cr
& & +(q-1)^2(q^6-10q^5+43q^4-102q^3+144q^2-120q+49)\xi \cr\cr
& & -(q-2)^2(q-1)^4=0 \ . 
\label{eqcub2}
\eeqs

The corresponding coefficients for the cyclic strip are
\beq
c_{hc3c,1}=c^{(3)}
\label{c1}
\eeq
\beq
c_{hc3c,j}=c^{(2)} \quad {\rm for} \ \ 2 \le j \le 5
\label{c25}
\eeq
\beq
c_{hc3c,j}=c^{(1)} \quad {\rm for} \ \ 6 \le j \le 11
\label{c611}
\eeq
\beq
c_{hc3c,j}=c^{(0)}=1 \quad {\rm for} \ \ 12 \le j \le 14 \ .
\label{c1214}
\eeq
For the M\"obius strip, as will be discussed further below in section 
\ref{cg}, the coefficients
are not, in general, simply $\pm 1$ times Chebyshev polynomials of the second
kind, $c^{(d)}$, but instead are products of the $c^{(d)}$ with certain terms
$\lambda_{cghc3,j}$'s occurring in the chromatic polynomial for related 
families of strip graphs called crossing-subgraph strips, defined via eqs. 
(\ref{lamcghc_12})-(\ref{eqcub3cg}).  We have 
\beq
c_{hc3mb,1}=-c^{(2)}=-(q^2-3q+1)
\label{c1mb}
\eeq
\beq
c_{hc3mb,2}=-c^{(0)}\lambda_{cghc3,3}=-(q-1)
\label{c2mb}
\eeq
\beq
c_{hc3mb,j}=-c^{(0)}\lambda_{cghc3,j+2} \quad {\rm for} \ \ j=3,4,5
\label{c25mb}
\eeq
\beq
c_{hc3mb,j}=c^{(1)}\lambda_{cghc3,j+5} \quad {\rm for} \ \ 6 \le j \le 11
\label{cmb611}
\eeq
\beq
c_{hc3mb,j}=c^{(0)}\lambda_{cghc3,j+5} \quad {\rm for} \ \ 12 \le j \le 14 \ .
\label{cmb1214}
\eeq

By general coloring matrix methods, we have
\beq
C_{P,G}=q\Bigl (q(q-1) \Bigr )^{\frac{L_y-1}{2}} 
\quad {\rm for} \quad G=hc, L_y
\times m,FBC_y,PBC_x, \ \ L_y \ \ {\rm odd}
\label{csumhclyodd}
\eeq
\beq
C_{P,G}=\Bigl (q(q-1) \Bigr )^{\frac{L_y}{2}} \quad {\rm for} \quad G=hc, L_y
\times m,FBC_y,PBC_x, \ \ L_y \ \ {\rm even}
\label{csumhclyeven}
\eeq

Our results for the M\"obius strips lead to the inference that 
\beq
C_{P,G}=0 \quad {\rm for} \quad G=hc, L_y \times m,FBC_y,TPBC_x \ .
\label{csumhcmbly3}
\eeq
The first special case of this, for $L_y=2$, was obtained in \cite{pg}. 

\begin{figure}[hbtp]
\centering
\leavevmode
\epsfxsize=2.5in
\begin{center}
\leavevmode
\epsffile{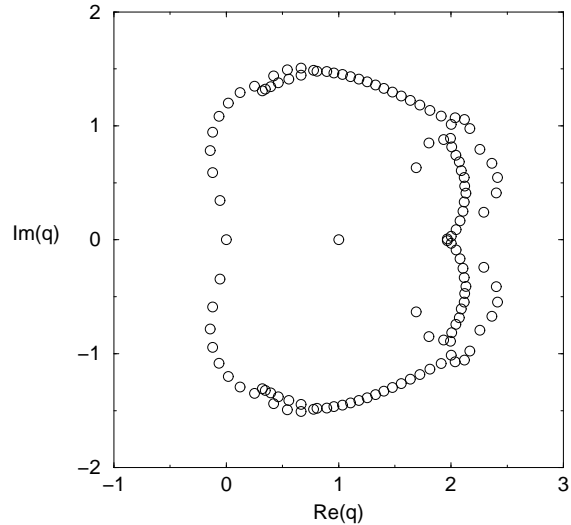}
\end{center}
\caption{\footnotesize{Chromatic zeros for the $L_y=3$ cyclic strip of the
honeycomb lattice of length $m=20$, i.e., $n=120$ vertices.}}
\label{hpxy3}
\end{figure}

As a special case of the arguments given in \cite{pm}, since the
$\lambda_{G,j}$'s are the same for the cyclic and M\"obius strips, and since in
the limit $m \to \infty$ the continuous singular locus ${\cal B}$ is defined as
the solution to the degeneracy of leading terms $\lambda_{G,j}$, it follows
that this locus is the same for the cyclic and M\"obius strips.  
In Fig. \ref{hpxy3} we show the chromatic zeros for a cyclic strip
of the honeycomb lattice with width $L_y=3$ and length, $m=20$, i.e., 
$n=120$ vertices.  This length is sufficiently great that the chromatic zeros 
for the cyclic and M\"obius strips are generally similar and lie close to the 
asymptotic singular locus ${\cal B}$.  We find the exact result 
\beq
q_c =2 \quad {\rm for} \quad hc, 3 \times \infty, FBC_y,(T)PBC_x \ .
\label{qchcyclic}
\eeq 
This is the same value that was obtained in \cite{pg} for the ($m \to \infty$
limit of the) $L_y=2$ cyclic or M\"obius strip of the honeycomb lattice (below
we shall show that the $L_y=4$ strip with cylindrical boundary conditions
yields a value of $q_c$ closer to the value (\ref{qchc}) for the 
infinite 2D honeycomb lattice).  The locus ${\cal B}$ for $L_y=3$ has support
for $Re(q) < 0$ as well as $Re(q) > 0$, just as was true for $L_y=2$
\cite{pg}.   

This $L_y=3$ locus separates the $q$ plane into several regions.  The outermost
one, region $R_1$, extends to infinite $|q|$ and includes the intervals $q >
2$ and $q < 0$ on the real $q$ axis.  Region $R_2$ includes the real interval
$0 < q < 2$.  Two complex-conjugate regions $R_3,R_3^*$, centered at
approximately $q=2.25 \pm 0.55i$, occur to the upper and lower right of $q_c$,
as is evident in Fig. \ref{hpxy3}.  Two further complex-conjugate regions,
$R_4,R_4^*$, centered at approximately $q=1.9 \pm 0.5i$, occur to the upper and
lower left of $q_c$.  One can observe that, just as was the case for $L_y=2$,
the density of zeros is small on the boundary separating region $R_4$ from
$R_2$ and its complex conjugate boundary separating $R_4^*$ from $R_2$.  As
previous exact calculations have shown \cite{strip2,wcy,t}, region diagrams for
infinite-length, finite-width strips of lattices can also include extremely
small regions. We have found evidence for a complex-conjugate pair of these
tiny regions to the upper and lower right of $q_c$.  We have not made an
exhaustive search for other tiny regions.

In region $R_1$, the dominant $\lambda_{hc3,j}$ is the root of the cubic
equation (\ref{eqcub2}) with the largest magnitude, which we denote as 
$\lambda_{hc3,12}$. Hence, 
\beq
W=(\lambda_{hc3,12})^{1/6} \ , \quad q \in R_1 \ .
\label{whc3_r1}
\eeq
The fact that this is the same as $W$ for the (FBC$_y$,FBC$_x$) case 
\cite{strip} is a general result \cite{bcc}.  
In region $R_2$, and also in regions $R_3,R_3^*$, the dominant 
terms are the roots of eq. (\ref{eqsix}) that are largest in magnitude in these
respective regions.  We shall refer to these respective dominant terms 
generically as $\lambda_{hc3,6}$, so that 
\beq
|W| = |\lambda_{hc3,6}|^{1/6} \ , \quad q \in R_2, R_3, R_3^* \ .
\label{whc3_r2}
\eeq
In regions $R_4, R_4^*$, the dominant term is the root of eq.
(\ref{eqcub1})
with maximal magnitude, which we denote as $\lambda_{hc3,3}$, so that 
\beq
|W|=|\lambda_{hc3,3}|^{1/6} \ , \quad q \in R_4 \ , R_4^* \ .
\label{whc_r4}
\eeq
Note that $W=1$ at $q_c=2$. A summary of some features of these strip graphs
is given in Table \ref{proptable}.  Regarding the comparison with previous
results, those for the $L_y=2$ cyclic/M\"obius strips are from Ref. \cite{pg}
and those for the $L_y=3$ open strip are from Ref. \cite{strip}. 

\begin{table}[hbtp]
\caption{\footnotesize{Properties of $P$, $W$, and ${\cal B}$ for strip graphs
$G_s$ of the honeycomb (hc) lattice.  The properties apply
for a given strip of size $L_y \times m$; some apply for
arbitrary $m$, such as $N_\lambda$, while others apply for the
infinite-length limit, such as the properties of the locus ${\cal B}$. 
For the boundary conditions in the $y$ and $x$ directions ($BC_y$,
$BC_x$), F, P, and T denote free, periodic, and orientation-reversed (twisted)
periodic, and the notation (T)P means that the results apply for either
periodic or orientation-reversed periodic.  The column denoted eqs. describes
the numbers and degrees of the algebraic equations giving the
$\lambda_{G_s,j}$; for example, $\{2(1),2(3),1(6)\}$ indicates that there are 2
linear equations, 2 cubic equations and one sixth degree equation.  The column
denoted BCR lists the points at which ${\cal B}$ crosses the real $q$ axis and 
the value of $q_c$ for the given family. The notation ``none'' in
this column indicates that ${\cal  B}$ does not cross the real $q$ axis.  The 
notation ``int;$q_1,q_c$'' refers to cases where ${\cal B}$ contains a
real
interval, there is a crossing at $q_1$, and the right-hand endpoint of the
interval is $q_c$. Column labelled ``SN'' refers to whether ${\cal B}$ has
\underline{s}upport for \underline{n}egative $Re(q)$, indicated as yes (y) or
no (n).}}
\begin{center}
\begin{tabular}{|c|c|c|c|c|c|c|}
\hline\hline 
$L_y$ & $BC_y$ & $BC_x$ & $N_\lambda$ & eqs. & BCR & SN \\ \hline\hline
2 & F  & F &    1   & \{1(1)\}           & none         & n       \\ \hline
3 & F  & F &    3   & \{1(3)\}           & 2            & n       \\ \hline
4 & F  & F &    5   & \{1(5)\}           & int;2.093,2.099  & n  \\ \hline
2 & F  & (T)P & 4   & \{4(1)\}           & 0, \ 2       & y       \\ \hline
3 & F  & (T)P & 14  & \{2(1),2(3),1(6)\} & 0, \ 2       & y       \\ \hline
4 & P  & F    & 4   & \{1(4)\}           & int;2.222,2.250  & y \\ \hline\hline
\end{tabular}
\end{center}
\label{proptable}
\end{table}

\section{$L_y=3$ Cyclic Crossing-Subgraph Strips of the Honeycomb Lattice} 
\label{cg}

An illustration of the cyclic $L_y=3$ crossing-subgraph strip of the 
honeycomb lattice, denoted $cghc3$, of length $\ell=5$, is shown in Fig. 
\ref{hcstrips} (c).  For this type of strip, with arbitrarily great
$\ell$, we 
calculate $N_{cghc3,cyc.}=19$ and
\beq
P(cghc,3 \times \ell, FBC_y,PBC_x,q) = \sum_{j=1}^{19} c_{cghc3,j}
(\lambda_{cghc3,j})^\ell
\label{pcghc3cyc}
\eeq
where for $1 \le j \le 4$, 
\beq
\lambda_{cghc3,1}=-\lambda_{cghc3,2}=1
\label{lamcghc_12}
\eeq
\beq
\lambda_{cghc3,3}=-\lambda_{cghc3,4}=q-1 \ .
\label{lamcghc_34}
\eeq
The $\lambda_{cghc3,j}$ for $j=5,6,7$ are the roots of the cubic equation
\beq
\xi^3+\xi^2-(q^2-4q+5)\xi-(q-1)=0 \ .
\label{eqcub1cg}
\eeq
The $\lambda_{cghc3,j}$ for $j=8,9,10$ are the roots of the cubic equation
\beq
\xi^3-\xi^2-(q^2-4q+5)\xi+(q-1)=0 \ .
\label{eqcub2cg}
\eeq
Since eq. (\ref{eqcub2cg}) is related to eq. (\ref{eqcub1cg}) by the reversal
of the signs of the coefficients of the $\xi^2$ and $\xi^0$ terms, it follows
that the respective roots of these equations are opposite in sign:
\beq
\lambda_{cghc3,j}=-\lambda_{cghc3,j+3} \quad {\rm for} \ \ j=5,6,7 \ .
\label{lamcghc_510rel}
\eeq
The $\lambda_{cghc3,j}$ for $11 \le j \le 16$ are the roots of the sixth degree
equation
\beqs
& & \xi^6+(q^2-4q+5)\xi^5-(q^4-5q^3+10q^2-10q+6)\xi^4 \cr\cr
& & -(q^6-10q^5+43q^4-102q^3+142q^2-110q+38)\xi^3 \cr\cr
& & +(q-1)(q^6-10q^5+43q^4-100q^3+132q^2-94q+29)\xi^2 \cr\cr
& & +(q-1)^2(q^2-4q+5)\xi-(q-2)(q-1)^4=0 \ .
\label{eqsixcg}
\eeqs
Finally, the $\lambda_{cghc3,j}$ for $j=17,18,19$ are the roots of the cubic
equation 
\beq
\xi^3-(q-2)(q^2-2q+3)\xi^2+(q-1)(q^3-5q^2+8q-5)\xi-(q-2)(q-1)^2=0 \ .    
\label{eqcub3cg}
\eeq

The corresponding coefficients are
\beq
c_{cghc3,1}=\frac{1}{2}(c^{(3)}-c^{(2)})=\frac{1}{2}(q-2)(q^2-4q+1)
\label{ccghc3_1}
\eeq
\beq
c_{cghc3,2}=\frac{1}{2}(c^{(3)}+c^{(2)})=\frac{1}{2}q(q-1)(q-3)
\label{ccghc3_2}
\eeq
\beq
c_{cghc3,j}=\frac{1}{2}(c^{(2)}-c^{(0)}) = \frac{1}{2}q(q-3) 
\quad {\rm for} \ \ j=3,5,6,7
\label{ccghc3_3}
\eeq
\beq
c_{cghc3,j}=\frac{1}{2}(c^{(2)}+c^{(0)}) = \frac{1}{2}(q-1)(q-2) 
\quad {\rm for} \ \ j=4,8,9,10
\label{ccghc3_4}
\eeq
\beq
c_{cghc3,j}=q-1 \quad {\rm for} \ \ 11 \le j \le 16
\label{ccghc3_11}
\eeq
\beq
c_{cghc3,j}=1 \quad {\rm for} \ \ 17 \le j \le 19 \ .
\label{ccghc3_17}
\eeq

In \cite{tor4} we have discussed how cyclic crossing subgraph strips of the
square lattice with free transverse boundary conditions reduce, for even and
odd length $L_x$, to cyclic and M\"obius strips, respectively.  Similarly,
here, for even and odd length $\ell$, the cyclic crossing subgraph strip of the
honeycomb lattice reduces to the cyclic and M\"obius strips of this lattice, so
that the chromatic polynomial for this cyclic crossing subgraph strip reduces
to the respective chromatic polynomial of the cyclic and M\"obius strip.  With
our labelling conventions for the length of the cyclic crossing subgraph strips
as compared with the cyclic or M\"obius strips of the honeycomb lattice, the
cyclic crossing subgraph of even length $\ell=2m$ (odd length $\ell=2m+1$) is
identical to the cyclic (M\"obius) strip of this lattice of length $m$,
respectively.  There is a resultant reduction in the number of
$\lambda_{G,j}$'s, from 19 to 14, as a result of the fact that there are five
pairs of terms $\lambda_{cghc3,j}$ that differ in sign and have different
coefficients, as given above.  For the even-length case we
thus have the identifications 
\beq 
(\lambda_{cghc3,j})^{2m}=(\lambda_{hc3,j})^m
\label{lamsqeven}
\eeq
while for the odd-length case, 
\beq
(\lambda_{cghc3,j})^{2m+1}=\lambda_{cghc3,j}(\lambda_{hc3,j})^m
\label{lamsqodd}
\eeq
so that, for both the even- and odd-$\ell$ cases for the crossing-subgraph
strips,
i.e., for both the cyclic and M\"obius strips, 
\beq
\lambda_{hc3,1}=(\lambda_{cghc3,j})^2 \quad {\rm for} \ \ j=1,2
\label{lamrel1}
\eeq
\beq
\lambda_{hc3,2}=(\lambda_{cghc3,j})^2 \quad {\rm for} \ \ j=3,4
\label{lamrel2}
\eeq
\beq
\lambda_{hc3,3}=(\lambda_{cghc3,j})^2 \quad {\rm for} \ \ j=5,8
\label{lamrel3}
\eeq
\beq
\lambda_{hc3,4}=(\lambda_{cghc3,j})^2 \quad {\rm for} \ \ j=6,9
\label{lamrel4}
\eeq
\beq
\lambda_{hc3,5}=(\lambda_{cghc3,j})^2 \quad {\rm for} \ \ j=7,10
\label{lamrel5}
\eeq
\beq
\lambda_{hc3,j}=(\lambda_{cghc3,j+5})^2 \quad {\rm for} \ \ 6 \le j \le 14
\ .
\label{lamrel614}
\eeq
In the case of odd length, we absorb the left-over factor of
$\lambda_{cghc3,j}$ in eq. (\ref{lamsqodd}) in the definition of the
coefficients.  
The relations (\ref{lamrel1})-(\ref{lamrel614}) expressing 
$\lambda_{hc3,j}$'s for the cyclic/M\"obius strips as squares of 
$\lambda_{cghc3,k}$'s are reflected in obvious relations between the algebraic
equations for these terms.

In the limit $\ell \to \infty$, one derives the singular locus ${\cal
B}$ for the crossing-subgraph strip.  As discussed in \cite{tor4}, since one
can take this limit on even values of $\ell$ or, alternatively, on odd values
of $\ell$, it follows that the singular loci ${\cal B}$ for the infinite-length
limits of the cyclic and M\"obius strips of the honeycomb lattice are the
same.  Here this already follows from the property that the $\lambda_{G,j}$'s 
are the same for the cyclic and M\"obius strips.  

\section{Structural Properties of Chromatic Polynomials for Cyclic Strips of
the Honeycomb Strip} 

Let us define $n_P(hc,L_y,d)$ as the number of terms $\lambda_{P,hc,L_y,j}$,
each of which has, as its coefficient, the Chebyshev polynomial $c^{(d)}$, in
the chromatic polynomial $P(hc,L_y,q)$ of the cyclic strip of the honeycomb
lattice of width $L_y$, denoted $hc,L_y$.  Clearly, the total number,
$N_{P,hc,L_y,\lambda}$, of different terms $\lambda_{P,hc,L_y,j}$ in
(\ref{pgsum}) for this family of strip graphs is given by 
\beq
N_{P,hc,L_y,\lambda} = \sum_{d=0}^{L_y} n_P(hc,L_y,d) \ .
\label{npsum}
\eeq 
Using the general formulas (\ref{csumhclyodd}) and (\ref{csumhclyeven})
together with the coloring matrix methods that we have discussed in detail in
\cite{cf}, we can calculate the numbers $n_P(hc,L_y,d)$ and
$N_{P,hc,L_y,\lambda}$.  We find
\beq
n_P(hc,L_y,d)=0 \quad {\rm for} \quad d > L_y \ ,
\label{npup}
\eeq
\beq
n_P(hc,L_y,L_y)=1
\label{npcly}
\eeq
and the values $n_P(hc,2,0)=1$, $n_P(hc,2,1)=2$.  All other numbers
$n_P(hc,L_y,d)$ are then determined by the following recursion relations.
For even $L_y \ge 4$
\beq
n_P(hc,L_y,d)=n_P(hc,L_y-1,d-1)+n_P(hc,L_y-1,d)+n_P(hc,L_y-1,d+1)
\quad {\rm for} \ \ d \ge 1
\label{nprecursionlyeven}
\eeq
while for $d=0$, 
\beq
n_P(hc,L_y,0)=n_P(hc,L_y-1,1) \ . 
\label{nprecursiond0lyeven}
\eeq
For odd $L_y \ge 3$
\beq
n_P(hc,L_y,d)=n_P(hc,L_y-1,d-1)+2n_P(hc,L_y-1,d)+n_P(hc,L_y-1,d+1) 
\quad {\rm for} \ \ d \ge 1
\label{nprecursionlyodd}
\eeq
while for $d=0$, 
\beq
n_P(hc,L_y,0)=n_P(hc,L_y-1,0)+n_P(hc,L_y-1,1) \ . 
\label{nprecursiond0lyodd}
\eeq
We observe that the recursion relations (\ref{nprecursionlyeven}) and 
(\ref{nprecursiond0lyeven}) are the same as we found in \cite{cf} for 
the numbers $n_P(L_y,d)$ for cyclic
strips of the square and triangular lattices, while eqs. 
(\ref{nprecursionlyodd}) and (\ref{nprecursiond0lyodd}) are analogous to the
recursion relations that we found for the numbers $n_Z(L_y,d)$ for cyclic
strips of the square and triangular lattices (in \cite{cf} these applied for
both even and odd $L_y$).  We also obtain the relations 
\beq
N_{P,hc,L_y,\lambda}=4N_{P,hc,L_y-1,\lambda}-2n_P(hc,L_y-1,0) \quad {\rm for \
odd} \ \ L_y \ge 3
\label{nptotlyodd}
\eeq
and
\beq
N_{P,hc,L_y,\lambda}=3N_{P,hc,L_y-1,\lambda}-2n_P(hc,L_y-1,0) 
\quad {\rm for \ even} \ \ L_y \ge 4
\label{nptotlyeven}
\eeq
whence
\beq
N_{P,hc,L_y,\lambda}=12N_{P,hc,L_y-2,\lambda}-8n_P(hc,L_y-2,0)-2n_P(hc,L_y-2,1)
\quad {\rm for} \ \ L_y \ge 4 \ .
\label{nptotlyrel}
\eeq
Results for $L_y$ up to 10 are given in Table \ref{nptable}.  These may be
compared with the analogous calculations that we carried out for cyclic strips
of the square and triangular lattices in \cite{cf}.  For example, the values 
$N_{P,hc,L_y,\lambda}=4,14$ and 36 for $L_y=2,3$ and 4 may be contrasted with 
the numbers $N_{P,G,L_y,\lambda}=4,10$ and 26 for $G=sq, \ tri$ with the same 
set of $L_y$ values.  It is straightforward to extend these to
higher values of $L_y$, and related results can also be given for M\"obius 
strips of the honeycomb lattice.

\begin{table}[hbtp]
\caption{\footnotesize{Table of numbers $n_P(hc,L_y,d)$ and their sums,
$N_{P,hc,L_y,\lambda}$ for cyclic strips of the honeycomb lattice. 
Blank entries are zero.}}
\begin{center}
\begin{tabular}{|c|c|c|c|c|c|c|c|c|c|c|c|c|}
\hline\hline
$L_y \ \downarrow$ \ \ $d \ \rightarrow$
   & 0 & 1   & 2   & 3   & 4   & 5  & 6  & 7 & 8 & 9 & 10 &
$N_{P,hc,L_y,\lambda}$
\\ \hline\hline
2  & 1   & 2   & 1   &     &     &    &    &   &   &    &   & 4    \\ \hline
3  & 3   & 6   & 4   & 1   &     &    &    &   &   &    &   & 14   \\ \hline
4  & 6   & 13  & 11  & 5   & 1   &    &    &   &   &    &   & 36   \\ \hline
5  & 19  & 43  & 40  & 22  & 7   & 1  &    &   &   &    &   & 132   \\ \hline
6  & 43  & 102 & 105 & 69  & 30  & 8  & 1  &   &   &    &   & 358  \\ \hline
7  & 145 & 352 & 381 & 273 & 137 & 47 & 10 & 1 &   &    &   & 1346  \\ \hline
8  & 352 & 878 & 1006& 791 & 457 & 194& 58 & 11& 1 &    &   & 3748  \\ \hline
9  & 1230& 3114& 3681& 3045& 1899& 903& 321& 81& 13& 1  &   & 14288 \\ \hline
10 & 3114& 8025& 9840& 8625& 5847& 3123&1305&415&95& 14 & 1 & 40404 \\
\hline\hline
\end{tabular}
\end{center}
\label{nptable}
\end{table}

  From eq. (\ref{nptotlyrel}), it follows that
the leading asymptotic behavior of $N_{P,hc,L_y,\lambda}$ (ignoring power-law
prefactors) is
\beq
N_{P,hc,L_y,\lambda} \sim (12)^{L_y/2} = (3.4641...)^{L_y} 
\quad {\rm as} \ \ L_y \to \infty \ .
\label{nptotlyasymp}
\eeq
This may be contrasted with the leading asymptotic behavior of the
corresponding numbers for the cyclic/M\"obius strips of the square and
triangular lattices \cite{cf} 
\beq
N_{P,G,L_y,\lambda} \sim 3^{L_y} \quad {\rm for} \ \ G=sq, \ tri 
\quad {\rm as} \ \ L_y \to \infty \ .
\label{nptotlyasympsqtri}
\eeq

Regarding the full Potts model partition function, we note that a
generalization of the relation (\ref{czsumcyc}) to $C_Z=q^{L_y}$ for the cyclic
honeycomb strip with width $L_y$ yields the result $N_{Z,hc,L_y,\lambda}=
{ 2L_y \choose L_y}$, similarly to the case with the corresponding strips of
the square and triangular lattices.

\section{Cylindrical $L_y=4$ Strip of the Honeycomb Lattice}

\begin{figure}[hbtp]
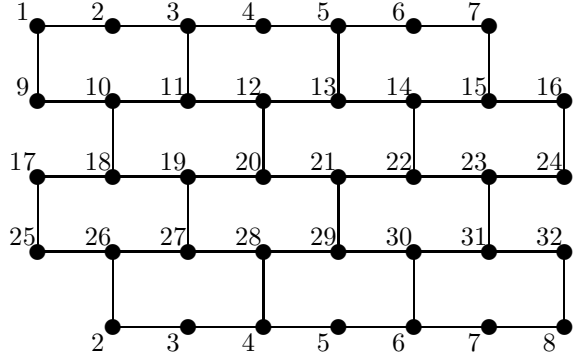

\caption{\footnotesize{Illustrative strip graph of the honeycomb lattice with
width $L_y=4$, length $m=3$ bricks and $(PBC_y,FBC_x)=$ cylindrical boundary 
conditions.}}
\label{cylstrip}
\end{figure}

The first width for which a strip of the honeycomb lattice with cylindrical
$=(PBC_y,FBC_x)$ boundary conditions can naturally be defined, satisfying
$\Delta_{max}=3$, is the width $L_y=4$.  A picture of
such a lattice is shown in Fig. \ref{cylstrip} for length $m=3$ bricks. In the
labelling convention of \cite{strip,hs}, this would be denoted as $m=2$; the
present convention yields somewhat simpler expressions for the coefficient
functions in the numerator of the generating function. 

\vspace*{1cm}
\begin{center}
\begin{picture}(70,40)
\multiput(10,0)(10,0){7}{\circle*{2}}
\multiput(0,10)(10,0){8}{\circle*{2}}
\multiput(0,20)(10,0){8}{\circle*{2}}
\multiput(0,30)(10,0){8}{\circle*{2}}
\multiput(0,40)(10,0){7}{\circle*{2}}
\multiput(10,0)(20,0){4}{\line(0,1){10}}
\multiput(0,10)(20,0){4}{\line(0,1){10}}
\multiput(10,20)(20,0){4}{\line(0,1){10}}
\multiput(0,30)(20,0){4}{\line(0,1){10}}
\put(10,0){\line(1,0){60}}
\multiput(0,10)(0,10){3}{\line(1,0){70}}
\put(0,40){\line(1,0){60}}
\put(8,-2){\makebox(0,0){2}}
\put(18,-2){\makebox(0,0){3}}
\put(28,-2){\makebox(0,0){4}}
\put(38,-2){\makebox(0,0){5}}
\put(48,-2){\makebox(0,0){6}}
\put(58,-2){\makebox(0,0){7}}
\put(68,-2){\makebox(0,0){8}}
\put(-2,12){\makebox(0,0){25}}
\put(8,12){\makebox(0,0){26}}
\put(18,12){\makebox(0,0){27}}
\put(28,12){\makebox(0,0){28}}
\put(38,12){\makebox(0,0){29}}
\put(48,12){\makebox(0,0){30}}
\put(58,12){\makebox(0,0){31}}
\put(68,12){\makebox(0,0){32}}
\put(-2,22){\makebox(0,0){17}}
\put(8,22){\makebox(0,0){18}}
\put(18,22){\makebox(0,0){19}}
\put(28,22){\makebox(0,0){20}}
\put(38,22){\makebox(0,0){21}}
\put(48,22){\makebox(0,0){22}}
\put(58,22){\makebox(0,0){23}}
\put(68,22){\makebox(0,0){24}}
\put(-2,32){\makebox(0,0){9}}
\put(8,32){\makebox(0,0){10}}
\put(18,32){\makebox(0,0){11}}
\put(28,32){\makebox(0,0){12}}
\put(38,32){\makebox(0,0){13}}
\put(48,32){\makebox(0,0){14}}
\put(58,32){\makebox(0,0){15}}
\put(68,32){\makebox(0,0){16}}
\put(-2,42){\makebox(0,0){1}}
\put(8,42){\makebox(0,0){2}}
\put(18,42){\makebox(0,0){3}}
\put(28,42){\makebox(0,0){4}}
\put(38,42){\makebox(0,0){5}}
\put(48,42){\makebox(0,0){6}}
\put(58,42){\makebox(0,0){7}}
\end{picture}
\end{center}

In general, for the strip of the honeycomb lattice with cylindrical
$=(PBC_y,FBC_x)$ boundary
conditions there are $n=2L_y(m+1)$ vertices and $e=L_y(3m+2)$ edges.
As before, it is
convenient to present the results in terms of a generating function, denoted
$\Gamma(G_s,q,x)$, where $G_s$ refers to the type of strip graph.  
The chromatic polynomial $P((G_s)_m,q)$ is determined as
the coefficient in a Taylor series expansion of this generating function in an
auxiliary variable $z$ about $z=0$:
\beq
\Gamma(G_s,q,z) =
\sum_{m=0}^{\infty}P((G_s)_m,q)z^m \ .
\label{gamma}
\eeq
The generating function $\Gamma(G_s,q,z)$ is a rational function of the form
\beq
\Gamma(G_s,q,z) = \frac{{\cal N}(G_s,q,z)}{{\cal D}(G_s,q,z)}
\label{gammagen}
\eeq
with
\beq
{\cal N}(G_s,q,z) = \sum_{j=0}^{d_{\cal N}} A_{G_s,j}(q) z^j
\label{n}
\eeq
and
\beq
{\cal D}(G_s,q,z) = 1 + \sum_{j=1}^{d_{\cal D}} b_{G_s,j}(q) z^j
\label{d}
\eeq
where the $A_{G_s,j}$ and $b_{G_s,j}$ are polynomials in $q$, and
$d_{\cal N} \equiv deg_z({\cal N})$, $d_{\cal D} \equiv deg_z({\cal D})$,
In factorized form
\beq
{\cal D}(G_s,q,z) = \prod_{j=1}^{d_{\cal D}}(1-\lambda_{G_s,j}(q)z) \ .
\label{lambdaform}
\eeq

We find that $deg_z({\cal N})=3$ and $deg_z({\cal D})=N_{\lambda}=4$. The
coefficient functions in the numerator and denominator of the generating
function are
\beq
A_0=q(q-1)(q^6-7q^5+21q^4-35q^3+35q^2-21q+7)
\label{a0hcly4cyl}
\eeq
\beqs
A_1 & = & -q(q-1)(q^{10}-15q^9+103q^8-417q^7+1074q^6-1782q^5+1850q^4 \cr\cr
& & -1125q^3+422q^2-246q+161)
\label{a1hcly4cyl}
\eeqs
\beqs
A_2 & = & q(q-1)(4q^{10}-64q^9+460q^8-1956q^7+5461q^6-10497q^5 \cr\cr
& & +14128q^4-13212q^3+8281q^2-3203q+614)
\label{a2hcly4cyl}
\eeqs
\beq
A_3=-4q(q-1)^5(q-2)^2
\label{a3hcly4cyl}
\eeq
\beq
b_1=-q^8+12q^7-66q^6+220q^5-496q^4+796q^3-922q^2+734q-314
\label{b1hcly4cyl}
\eeq
\beqs
b_2 & = & q^{12}-20q^{11}+188q^{10}-1092q^9+4344q^8-12428q^7+26192q^6 \cr\cr
& & -41022q^5+47597q^4-40318q^3+24247q^2-9782q+2169
\label{b2hcly4cyl}
\eeqs
\beqs
b_3 & = & -4q^{12}+80q^{11}-736q^{10}+4124q^9-15705q^8+42922q^7-86535q^6 
\cr\cr & & +129964q^5-144575q^4+116374q^3-64525q^2+22280q-3680 
\label{b3hcly4cyl}
\eeqs
\beq
b_4=4[(q-1)(q-2)]^4 \ .
\label{b4hcly4cyl}
\eeq
The $W$ function is given as $W=(\lambda_{hc4pf,j})^{1/8}$, where
$\lambda_{hc4pf,j}$ is the maximum-magnitude root of the equation
\beq
\xi^4 + b_1 \xi^3 + b_2 \xi^2 + b_3 \xi + b_4 = 0
\label{xieqpf}
\eeq
with the $b_j$ coefficients given in
eqs. (\ref{b1hcly4cyl})-(\ref{b4hcly4cyl}).

\begin{figure}[hbtp]
\centering
\leavevmode
\epsfxsize=2.5in
\begin{center}
\leavevmode
\epsffile{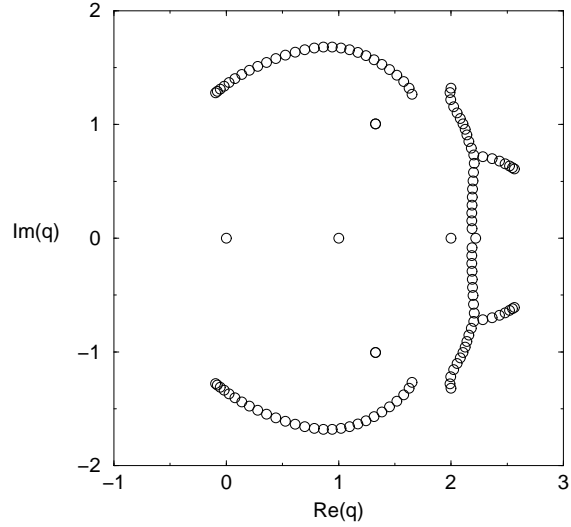}
\end{center}
\caption{\footnotesize{Chromatic zeros for the width $L_y=4$ strip of the 
honeycomb lattice with cylindrical boundary conditions, $(PBC_y,FBC_x)$, for
length $m=14$ (i.e., $n=120$).}}
\label{hpy4}
\end{figure}

A plot of the chromatic zeros for a long finite strip of this type is shown in
Fig. \ref{hpy4}.  For this great a length, these zeros give a reasonably
accurate indication of the position of the asymptotic locus ${\cal B}$. From
our exact analytic results, we find that the locus ${\cal B}$ includes arcs and
a short interval on the real axis $2.194004 \le q \le q_c$, where
$q_c=2.249840$.  It is interesting that this value of $q_c$ is just 14 \% below
the value (\ref{qchc}) for the infinite 2D
honeycomb lattice.  We also note that the locus ${\cal B}$ includes support for
$Re(q) < 0$, as is evident from Fig. \ref{hpy4}.

\section{Free $L_y=4$ Strip of the Honeycomb Lattice}

 In general for a strip of the honeycomb lattice with free boundary conditions,
width $L_y$, and $m$ bricks, there are $n=2(L_ym+L_y-1)$ vertices and
$e=(3L_y-1)m+2L_y-3$ edges.  An illustration of the strip of the honeycomb
lattice with width $L_y=4$, length $m=3$, and free $=(FBC_y,FBC_x)$ boundary
conditions is shown in Fig.  \ref{openstrip} (in the labelling convention of
\cite{strip,hs}, this would be denoted as length $m=2$).

\vspace*{1cm}
\begin{center}
\begin{picture}(70,30)
\multiput(10,0)(10,0){7}{\circle*{2}}
\multiput(0,10)(10,0){8}{\circle*{2}}
\multiput(0,20)(10,0){8}{\circle*{2}}
\multiput(10,30)(10,0){7}{\circle*{2}}
\multiput(10,0)(20,0){4}{\line(0,1){10}}
\multiput(0,10)(20,0){4}{\line(0,1){10}}
\multiput(10,20)(20,0){4}{\line(0,1){10}}
\put(10,0){\line(1,0){60}}
\multiput(0,10)(0,10){2}{\line(1,0){70}}
\put(10,30){\line(1,0){60}}
\put(8,-2){\makebox(0,0){24}}
\put(18,-2){\makebox(0,0){25}}
\put(28,-2){\makebox(0,0){26}}
\put(38,-2){\makebox(0,0){27}}
\put(48,-2){\makebox(0,0){28}}
\put(58,-2){\makebox(0,0){29}}
\put(68,-2){\makebox(0,0){30}}
\put(-2,12){\makebox(0,0){16}}
\put(8,12){\makebox(0,0){17}}
\put(18,12){\makebox(0,0){18}}
\put(28,12){\makebox(0,0){19}}
\put(38,12){\makebox(0,0){20}}
\put(48,12){\makebox(0,0){21}}
\put(58,12){\makebox(0,0){22}}
\put(68,12){\makebox(0,0){23}}
\put(-2,22){\makebox(0,0){8}}
\put(8,22){\makebox(0,0){9}}
\put(18,22){\makebox(0,0){10}}
\put(28,22){\makebox(0,0){11}}
\put(38,22){\makebox(0,0){12}}
\put(48,22){\makebox(0,0){13}}
\put(58,22){\makebox(0,0){14}}
\put(68,22){\makebox(0,0){15}}
\put(8,32){\makebox(0,0){1}}
\put(18,32){\makebox(0,0){2}}
\put(28,32){\makebox(0,0){3}}
\put(38,32){\makebox(0,0){4}}
\put(48,32){\makebox(0,0){5}}
\put(58,32){\makebox(0,0){6}}
\put(68,32){\makebox(0,0){7}}
\end{picture}
\end{center}
\begin{figure}[hbtp]
\caption{\footnotesize{Illustrative strip graph of the honeycomb lattice with
width $L_y=4$, length $m=3$ bricks and open boundary conditions.}}
\label{openstrip}
\end{figure}

We find that $deg_x({\cal N})=4$ and $deg_x({\cal D})=N_{\lambda}=5$. The
coefficient functions in the numberator and denominator of the generating
function are
\beq
A_0=q(q-1)^5
\label{a0hcly4ff}
\eeq
\beq
A_1=-q(q-1)(2q^8-22q^7+109q^6-314q^5+566q^4-636q^3+417q^2-133q+10)
\label{a1hcly4ff}
\eeq
\beqs
A_2 & = & q(q-1)^2(q^{11}-17q^{10}+133q^9-628q^8+1975q^7-4306q^6+6570q^5 \cr\cr
& & -6933q^4+4921q^3-2273q^2+683q-133)
\label{a2hcly4ff}
\eeqs
\beqs
A_3 & = & -q(q-1)^2(2q^{11}-34q^{10}+268q^9-1289q^8+4182q^7-9547q^6 \cr\cr
& & +15548q^5-17985q^4+14487q^3-7811q^2+2599q-424)
\label{a3hcly4ff}
\eeqs
\beq
A_4=2q(q-1)^5(q-2)(2q^5-16q^4+49q^3-71q^2+49q-12)
\label{a4hcly4ff}
\eeq

\beq
b_1=-(q^2-4q+5)(q^6-7q^5+22q^4-42q^3+54q^2-47q+27)
\label{b1hcly4ff}
\eeq
\beqs
b_2 & = & 2q^{12}-36q^{11}+305q^{10}-1609q^9+5894q^8-15827q^7+32046q^6
\cr\cr & & -49513q^5+58255q^4-51198q^3+32069q^2-12915q+2555
\label{b2hcly4ff}
\eeqs
\beqs
b_3 & = & -q^{16}+25q^{15}-297q^{14}+2224q^{13}-11745q^{12}+46379q^{11} 
\cr\cr & & -141680q^{10}+341694q^9-658019q^8+1016383q^7-1256955q^6 \cr\cr & & 
+1233979q^5-945308q^4+548423q^3-228538q^2+61589q-8161 
\label{b3hcly4ff}
\eeqs
\beqs
b_4 & = & (q-1)^2(2q^{14}-44q^{13}+456q^{12}-2951q^{11}+13318q^{10}-44313q^9
\cr\cr & & +112043q^8-218692q^7+331398q^6-388712q^5+348419q^4	\cr\cr & & 
-232363q^3+109612q^2-33033q+4876)
\label{b4hcly4ff}
\eeqs
\beq
b_5 = -2(q-1)^6(q-2)^2(2q^6-20q^5+83q^4-185q^3+239q^2-175q+60) \ .
\label{b5hcly4ff}
\eeq
The $W$ function is given as $W=(\lambda_{hc4ff,j})^{1/8}$, where
$\lambda_{hc4ff,j}$ is the maximum-magnitude root of the equation
\beq
\xi^5 + b_1 \xi^4 + b_2 \xi^3 + b_3 \xi^2 + b_4 \xi + b_5= 0
\label{xieqff}
\eeq
with the $b_j$ coefficients given in eqs. (\ref{b1hcly4ff})-(\ref{b5hcly4ff}).

\begin{figure}[hbtp]
\centering
\leavevmode
\epsfxsize=2.5in
\begin{center}
\leavevmode
\epsffile{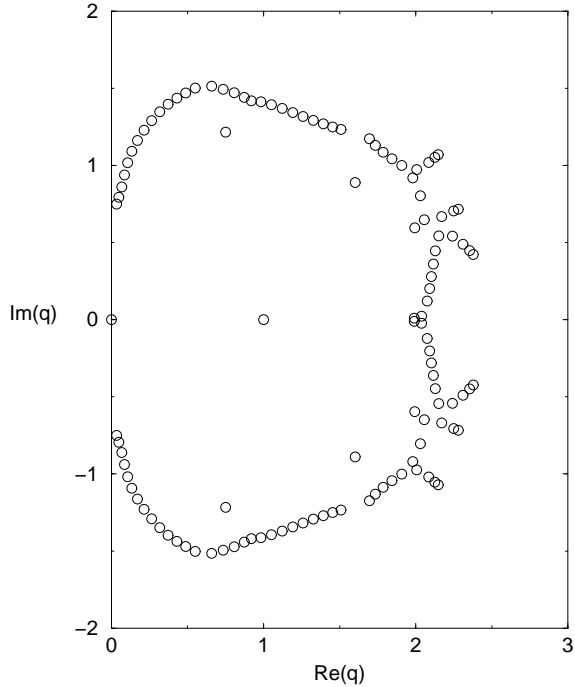}
\end{center}
\caption{\footnotesize{Chromatic zeros for the width $L_y=4$ strip of the
honeycomb lattice with free boundary conditions, $(FBC_y,FBC_x)$ and length
$m=14$ (i.e., $n=118$).}}
\label{hy4}
\end{figure}

A plot of the chromatic zeros for a long finite strip of this type is shown in
Fig. \ref{hy4}. From our exact analytic results we find that the locus 
${\cal B}$ includes arcs and a short interval on the real axis
$2.078191 \le q \le q_c$ where $q_c=2.099012$. 

\vspace{10mm}

\section{Numerical Values of $W$}

It is of interest to use our new exact results to study further the
approach of $W$ to the $L_y \to \infty$ limit, i.e., the infinite 2D 
honeycomb lattice.  This
extends our previous study in \cite{w2d}.  In Table \ref{hc2d} we list
values of $W(hc,L_y \times \infty, BC_y, BC_x,q)$ (which, for this
range of $q$, are independent of $BC_x$), for free and periodic (cylindrical) 
transverse boundary conditions.  These are compared with Monte Carlo 
measurements of $W$ for the full 2D square lattice, $W(hc,q)$ from 
\cite{ww,w3}. We also list the ratios 
\beq
R_W(\Lambda,L_y,BC_y,q) = \frac{W(\Lambda,L_y,BC_y,q)}{W(\Lambda,q)}
\label{rw}
\eeq 
for the present case, $\Lambda=hc$.  In \cite{w2d} it was proved for a
general lattice $\Lambda$ that the ratio (\ref{rw}) approaches unity
monotonically for $FBC_y$.  For $PBC_y$, this approach is nonmonotonic, but
more rapid.  One sees from Table \ref{hc2d} that for $L_y=4$ and moderate
values of $q$, say 5 or 6, the agreement of $W(hc,L_y,q)$ for the
infinite-length, finite-width open strips with the respective values $W(hc,q)$
for the infinite honeycomb lattice is very good; the ratios are unity to within
a few per cent.  For the same width $L_y=4$ and similar $q$ values, 
the agreement for the cylindrical strip is excellent; the ratios are
unity to within a few parts in $10^5$.  The agreement, as measured by these
ratios, gets better as $q$ increases.  This shows that calculations on
infinite-length, finite-width strips of the honeycomb lattice with cylindrical
boundary conditions provide an excellent approximation to the exact function
$W(hc,q)$ for the full 2D honeycomb lattice.  This is a very useful result
since the latter function has never been calculated exactly.  
We also give some further values of $W$ or $|W|$ in Table \ref{whcinternal}.

\begin{table}[hbtp]
\caption{\footnotesize{Values of $W(hc,L_y \times \infty,FBC_y,BC_x,q)$,
denoted $W(hc,L_y,F,q)$ for short, and 
$W(hc,L_y \times \infty,PBC_y,BC_x,q)$, denoted 
$W(hc,L_y,P,q)$ for short, with any $BC_x$.  These are compared with the
values for the full 2D honeycomb lattice, $W(hc,q)=W(hc,\infty
\times \infty,q)$ for $3 \le q \le 10$.  For each value of $q$, the quantities
in the upper line are identified at the top and the quantities in the lower
line are the values of $R_W(hc,L_y,BC_y,q)$.}}
\begin{center}
\begin{tabular}{|c|c|c|c|c|c|}
\hline\hline
$q$ & $W(hc,2,F,q)$ & $W(hc,3,F,q)$ & $W(hc,4,F,q)$ & $W(hc,4,P,q)$ & 
$W(hc,q)$ \\ \hline
3 &   1.82116 & 1.76567 & 1.73864  & 1.67079  & 1.6600(5)   \\
  &   1.097   & 1.064   & 1.047    & 1.0065   & 1           \\ \hline
4 &   2.79468 & 2.72942 & 2.69737  & 2.60443  & 2.6038(7)   \\
  &   1.073   & 1.048   & 1.036    & 1.0002   & 1           \\ \hline
5 &   3.78389 & 3.71448 & 3.68025  & 3.57963  & 3.5796(10)  \\
  &   1.057   & 1.038   & 1.028    & 1.0000   & 1           \\ \hline
6 &   4.77760 & 4.70568 & 4.67013  & 4.56513  & 4.5654(15)  \\
  &   1.046   & 1.031   & 1.023    & 0.9999   & 1           \\ \hline
7 &   5.77336 & 5.69973 & 5.66327  & 5.55529  & 5.5556(17)  \\
  &   1.039   & 1.026   & 1.019    & 0.9999   & 1           \\ \hline
8 &   6.77028 & 6.69539 & 6.65826  & 6.54810  & 6.5479(20)  \\
  &   1.034   & 1.0225  & 1.017    & 1.0000   & 1           \\ \hline
9 &   7.76793 & 7.69208 & 7.65443  & 7.54259  & 7.5424(22)  \\
  &   1.030   & 1.020   & 1.015    & 1.0000   & 1           \\ \hline
10 &  8.76607 & 8.68945 & 8.65140  & 8.53822  & 8.5386(25)  \\
  &   1.027   & 1.018   & 1.013    & 1.0000   & 1           \\ \hline\hline
\end{tabular}
\end{center}
\label{hc2d}
\end{table}

\begin{table}[hbtp]
\caption{\footnotesize{Values of $W(hc,L_y \times \infty,BC_y,BC_x,q)$ for low
integral $q$ and for respective $q_c$, if such a point exists, where $BC_y$ and
$BC_x$ denote the transverse and longitudinal boundary conditions. The notation
$n$ in the $q_c$ column means that there is no $q_c$ for this family, i.e., 
${\cal B}_q$ does not cross the positive real $q$ axis.}}
\begin{center}
\begin{tabular}{|c|c|c|c|c|c|c|c|}
\hline\hline
$L_y$ & $BC_y$ & $BC_x$ & $|W_{q=0}|$ & $|W_{q=1}|$ & $|W_{q=2}|$ & $q_c$ &
$W_{q=q_c}$ \\
\hline\hline
2  & F & F    & 1.495  & 1       & 1      & n      & $-$    \\ \hline
3  & F & F    & 1.70   & 1.26    & 1      & 2      & 1      \\ \hline
4  & F & F    & 1.81   & 1.39    & 1.09   & 2.10   & 1.06   \\ \hline\hline
2  & F & (T)P & 1.495  & 1.19    & 1      & 2      & 1      \\ \hline
3  & F & (T)P & 1.70   & 1.35    & 1      & 2      & 1      \\ \hline\hline
4  & P & F    & 2.05   & 1.56    & 1.19   & 2.25   & 1.12   \\ \hline\hline
\end{tabular}
\end{center}
\label{whcinternal}
\end{table}

\section{Conclusions} 

In this paper we have presented exact calculations of the partition function of
the $q$-state Potts model and the special case of the $T=0$ Potts
antiferromagnet (chromatic polynomial) on a variety of strips of the honeycomb
lattice with fixed width and arbitrarily great length.  We have computed the
exact free energy and discussed the thermodynamics for the ferromagnetic and
antiferromagnetic cases, including the singularities at the zero-temperature
critical point of the ferromagnet.  These results were compared with those
obtained from exact free energy calculations for infinite-length strips of the
square and triangular lattices and the differences explained.  The $W$ values
calculated from the finite-width, infinite-length strips were shown to approach
the values for the full 2D honeycomb lattice reasonably rapidly. This was
especially dramatic for the $L_y=4$ strip with periodic transverse
(cylindrical) boundary conditions, and the exact results for this strip thus
provide a very accurate approximation to the $W$ function for the full 2D
honeycomb lattice.  This is useful since there is no exact solution for $W(q)$
on the honeycomb lattice for $q > 2$.  Generalizing $q$ and temperature to
complex values and taking the limit of infinite length, we have studied the
locus of singularities ${\cal B}$ of the free energy in the $q$ and $u$ planes,
and for the $T=0$ antiferromagnet, the locus ${\cal B}_q$.

\vspace{10mm}

Acknowledgment: The research of R. S. was supported in part by the U. S. NSF
grant PHY-97-22101.

\section{Appendix}

\subsection{A Remark on Generating Functions}

In \cite{strip,strip2,hs} the methodology on generating function for chromatic
polynomials for recursive families of strip graphs was given.  For generality,
the open strip graphs were taken to be of the symbolic form 
$G_m=IH^m$, i.e., $m$-fold repetitions
of a subgraph $H$ connected to an initial end graph $I$.  It was proved in
\cite{strip} that the denominator of the generating function is independent of
the end graph $I$.  In \cite{strip2}, a study was carried out of inhomogeneous
strip graphs where $I \ne H$. However, if $I=H$, one can simplify the general
formalism slightly by redefining the label $m$ via a shift by one unit.  Thus,
if $I=H$, then the strip denoted symbolically as $IH^m$ is identical to
$H^{m+1}$.  The generating function is given in general by equations of the
form (\ref{gamma})-(\ref{d}). 
Now in the convention (which is denoted here as C1) of
\cite{strip,strip2,hs}, the zeroth-order term, $A_{G_s,0}$ in the expansion
(\ref{gamma}) is the chromatic polynomial for the subgraph $I$.  If $I=H$, 
as in the homogeneous strips considered here, 
then it is simpler to use a convention, which we denote C2, according to which
the zeroth-order term in the expansion is the chromatic polynomial for one end
(say the left-hand end) of the strip.  For example, for the width $L_y$ strip
of the square lattice, in conventions C1 and C2, $A_{sq,L_y,0}$ would be the
chromatic polynomial for a column of squares of height $L_y$ and the line 
graph $T_{L_y}$, respectively, i.e., $A_{sq,L_y,0}=q(q-1)(q^2-3q+3)^{L_y-1}$
and $A_{sq,L_y,0}=q(q-1)^{L_y-1}$.  To distinguish these two conventions, let
us, in this subsection, append a prime to the numerator defined with the 
latter convention C2, i.e., 
\beq
{\cal N}(G_s,q,z)^\prime = \sum_{j=0}^{d_{\cal N}} A_{G_s,j}^\prime(q) z^j \
.
\label{npc2}
\eeq
Thus, for $m \ge 1$, the $m$'th 
term in the expansion of the generating function 
$\Gamma(G_s,q,z)^\prime$ is the same as the $(m-1)$'th term in the expansion of
the generating function $\Gamma(G_s,q,z)$.  While the denominators are the
same, the coefficient functions in the numerators are related according to
\beq
A_{G_s,j}^\prime = A_{G_s,j-1} + A_{G_s,0}^\prime b_{G_s,j} 
\quad {\rm for} \ \ j \ge 1 \ .
\label{aarel}
\eeq
Similar relations hold for the generating functions of the full Potts model
partition function and equivalently the Tutte polynomial. 

We give some simple illustrations next.  For the $L_y=2$ strip of the square
lattice with free boundary conditions and C1, 
\beq
\Gamma_{P,sq,2} = \frac{q(q-1)D_4}{1-D_4z}
\label{gammasqly2openold}
\eeq
while with convention C2
\beq
\Gamma^\prime_{P,sq,2} = \frac{q(q-1)}{1-D_4z} \ .
\label{gammasqly2open}
\eeq
For the $L_y=3$ strip of the square lattice with free boundary conditions and
labelling convention C1 \cite{strip} 
\beq
A_{P,sq3,0}=q(q-1)D_4^2
\label{asq3p_0}
\eeq
\beq
A_{P,sq3,1}=-q(q-1)^3(q^3-6q^2+13q-11)
\label{asq3p_1}
\eeq
while for convention C2,
\beq
A_{P,sq3,0}^\prime =q(q-1)^2
\label{asq3pn_0}
\eeq
\beq
A_{P,sq3,1}^\prime =-q(q-1)(q^2-3q+1)
\label{asq3pn_1}
\eeq
and in both cases 
\beq
b_{P,sq3,1}=-(q-2)(q^2-3q+5)
\label{bsq3p_1}
\eeq
\beq
b_{P,sq3,2}=(q-1)(q^3-6q^2+13q-11) \ .
\label{bsq3p_2}
\eeq

Similarly, with convention C1, the generating functions for the Tutte 
polynomials 
for the $L_y=2$ open strip of the square and triangular lattices are
\beq
\Gamma_{T,sq,2}=\frac{(y+x+x^2+x^3)-yx^3z}{
1-(y+1+x+x^2)z+yx^2z^2}
\label{gammasq2t}
\eeq
and
\beq
\Gamma_{T,tri,2}=\frac{x(x+1)^2+2xy+y(y+1)-x^3y^2z}{
1-[(x+1)^2 + y(y+2)]z+x^2y^2z^2}
\label{gammatri2t}
\eeq
while with convention C2, they have the simpler forms 
\beq
\Gamma_{T,sq,2}^\prime =\frac{x+y(1-x)z}{1-(y+1+x+x^2)z+yx^2z^2}
\label{gammasq2tprime}
\eeq
and
\beq
\Gamma_{T,tri,2}^\prime =\frac{x-y(yx-y-1)z}{1-[(x+1)^2 +
y(y+2)]z+x^2y^2z^2}
\label{gammatri2tprime}
\eeq
and so forth for other lattice strips.  We shall use convention C2 here. 

\subsection{Connection Between Potts Model Partition Function and Tutte
Polynomial}

The formulas relating the Potts model partition function $Z(G,q,v)$ and the
Tutte polynomial $T(G,x,y)$ were given in \cite{a} and hence we shall be brief
here.  The Tutte polynomial of $G$, $T(G,x,y)$, is given by 
\cite{tutte1}-\cite{tutte3}
\beq
T(G,x,y)=\sum_{G^\prime \subseteq G} (x-1)^{k(G^\prime)-k(G)}
(y-1)^{c(G^\prime)}
\label{tuttepol}
\eeq
where $k(G^\prime)$, $e(G^\prime)$, and $n(G^\prime)=n(G)$ denote the number
of components, edges, and vertices of $G^\prime$, and
\beq
c(G^\prime) = e(G^\prime)+k(G^\prime)-n(G^\prime)
\label{ceq}
\eeq
is the number of independent circuits in $G^\prime$.
For the graphs of interest here, $k(G)=1$.  Now let
\beq
x=1+\frac{q}{v}
\label{xdef}
\eeq
and
\beq
y=a=v+1
\label{ydef}
\eeq
so that
\beq
q=(x-1)(y-1) \ .
\label{qxy}
\eeq
Then
\beq
Z(G,q,v)=(x-1)^{k(G)}(y-1)^{n(G)}T(G,x,y) \ .
\label{ztutte}
\eeq

For a planar graph $G$ the Tutte polynomial satisfies the duality relation
\beq
T(G,x,y) = T(G^*,y,x)
\label{tuttedual}
\eeq
where $G^*$ is the (planar) dual to $G$.  As discussed in \cite{a},
the Tutte polynomial for recursively defined graphs comprised of $m$
repetitions of some subgraph has the form
\beq
T(G_m,x,y) = \sum_{j=1}^{N_\lambda} c_{T,G,j}(\lambda_{T,G,j})^m \ .
\label{tgsum}
\eeq

\subsection{Open $L_y=2$ Strip of the Honeycomb Lattice}

The generating function for the Tutte polynomial for the open strip of the
honeycomb lattice comprised of $m$ bricks, denoted $S_m$, is \beq
\Gamma_T(S_m,x,y,z) = \sum_{m=0}^\infty T(S_m,x,y)z^m \ .
\label{gammatfbc}
\eeq
We calculate
\beq
\Gamma_T(S,x,y,z) = \frac{x+y(1-x)z}{1-(x^4+x^3+x^2+x+y+1)z+yx^4z^2} \ .
\label{gammats}
\eeq
This yields
\beq
\lambda_{T,S,(1,2)}=\frac{1}{2}\biggl [x^4+x^3+x^2+x+y+1 \pm \sqrt{R_{TS}} \ 
\biggr ]
\label{lamts12}
\eeq
where
\beqs
R_{TS} & = & x^8+2x^7+3x^6-2yx^4+4x^5+2yx^3+5x^4+2yx^2 \cr\cr 
& & +4x^3+y^2+2yx+3x^2+2y+2x+1 \ . 
\label{rts}
\eeqs

\subsection{Cyclic and M\"obius $L_y=2$ Strips of the Honeycomb Lattice}

We calculate 
\beq
T(L_m,x,y) = \sum_{j=1}^6 c_{T,L,j}(\lambda_{T,L,j})^m
\label{tlxy}
\eeq
and
\beq
T(ML_m,x,y) = \sum_{j=1}^6 c_{T,ML,j}(\lambda_{T,ML,j})^m
\label{tmbxy}
\eeq
where
\beq
\lambda_{T,ML,j}=\lambda_{T,L,j} \ , \quad j=1,...,6
\label{lamtutlmb}
\eeq
\beq
\lambda_{T,L,1} = 1
\label{lamtut1}
\eeq
\beq
\lambda_{T,L,2} = x^2
\label{lamtut2}
\eeq
\beq
\lambda_{T,L,(3,4)} = \frac{1}{2}\biggl [x^2+2x+y+2 \pm \sqrt{R_{T34}} \ 
\biggr ]
\label{lamtut34}
\eeq
with
\beq
R_{T34}=x^4+4x^3-2x^2y+8x^2+4xy+8x+y^2+4y+4
\label{rt34}
\eeq
\beq
\lambda_{T,L,5}=\lambda_{T,S,1} \ , \quad \lambda_{T,L,6}=\lambda_{T,S,2}
\label{lamtut56}
\eeq
where $\lambda_{T,S,1}$ and $\lambda_{T,S,2}$ were given above. 

It is convenient to extract a common factor from the coefficients:
\beq
c_{T,G,j} \equiv \frac{\bar c_{T,G,j}}{x-1} \ , \quad G = L, ML \ .
\label{cbar}
\eeq
Of course, although the individual terms contributing
to the Tutte polynomial are thus rational functions of $x$ rather than
polynomials in $x$, the full Tutte polynomial is a polynomial
in both $x$ and $y$.  We have
\beq
\bar c_{T,L,j}=c_{Z,L,j} \quad {\rm for} \ 1 \le j \le 6 \ .
\label{cj}
\eeq
Thus, in terms of the variables $x$ and $y$, e.g., $\bar c_{T,L,j}=q-1=
xy-x-y$ for $j=2,3,4$, etc. Further, 
\beq
\bar c_{T,ML,j}=c_{Z,ML,j} \quad {\rm for} \ 1 \le j \le 6 \ .
\label{cjml}
\eeq

\subsection{Special Values of Tutte Polynomials for Strips of the Honeycomb
Lattice}

For a given graph $G=(V,E)$, at certain special values of the arguments $x$ and
$y$, the Tutte polynomial $T(G,x,y)$ yields quantities of basic graph-theoretic
interest \cite{tutte3}-\cite{boll}.  We recall some definitions: a spanning
subgraph was defined at the beginning of the paper; a tree is a
connected graph with no cycles; a forest is a graph containing one or
more trees; and a spanning tree is a spanning subgraph that is a tree.  We
recall that the graphs $G$ that we consider are connected.  Then the number
of spanning trees of $G$, $N_{ST}(G)$, is
\beq
N_{ST}(G)=T(G,1,1) \ ,
\label{t11}
\eeq
the number of spanning forests of $G$, $N_{SF}(G)$, is
\beq
N_{SF}(G)=T(G,2,1) \ ,
\label{t21}
\eeq
the number of connected spanning subgraphs of $G$, $N_{CSSG}(G)$, is
\beq
N_{CSSG}(G)=T(G,1,2) \ ,
\label{T12}
\eeq
and the number of spanning subgraphs of $G$, $N_{SSG}(G)$, is
\beq
N_{SSG}(G)=T(G,2,2) \ .
\label{t22}
\eeq
An elementary theorem (e.g., \cite{ka}) is that 
\beq
N_{SSG}(G)=2^{e(G)} \ . 
\label{nssggen}
\eeq

  From our calculations of Tutte polynomials, we find that for the $L_y=2$
strip of the honeycomb lattice with free boundary conditions, $S_m$, 
\beq
N_{ST}(S_m)=\frac{1}{4\sqrt{2}} \biggl [ 
 (3+2\sqrt{2} \ )^{m+1}
-(3-2\sqrt{2} \ )^{m+1} \biggr ]
\label{t11sm}
\eeq

\beq
N_{SF}(S_m)=\frac{1}{8\sqrt{15}} \biggl [
(31+8\sqrt{15} \ )[4(4+\sqrt{15} \ )]^m-
(31-8\sqrt{15} \ )[4(4-\sqrt{15} \ )]^m \biggr ]
\label{t21sm}
\eeq

\beq
N_{CSSG}(S_m)=\frac{1}{\sqrt{41}} \Biggl [
 \biggl ( \frac{7+\sqrt{41}}{2} \biggr )^{m+1}
-\biggl ( \frac{7-\sqrt{41}}{2} \biggr )^{m+1} \Biggr ] 
\label{t12sm}
\eeq

\beq
N_{SSG}(S_m)=2^{e(S_m)}=2^{5m+1} \ .
\label{t22sm}
\eeq

For the $L_y=2$ cyclic and M\"obius strips of the honeycomb lattice, $L_m$ and
$ML_m$, we find, with $\eta_G=\pm 1$ for $L_m$ and $ML_m$, respectively:
\beq
N_{ST}(G_m)=m \biggl [-2\eta_G+(3+2 \sqrt{2} \ )^m+
                                       (3-2 \sqrt{2} \ )^m \biggr ] \quad 
{\rm for} \ \ G_m=L_m, \ ML_m
\label{t11lm}
\eeq

\beqs
N_{SF}(G_m) & = & \eta_G(1 - 2^{2m}) - \Biggl [ 
\biggl ( \frac{11+\sqrt{105} \ }{2} \biggr )^m +
\biggl ( \frac{11-\sqrt{105} \ }{2} \biggr )^m \Biggr ] \cr\cr & & + 
\Bigl \{ [4(4+\sqrt{15} \ )]^m + [4(4-\sqrt{15} \ )]^m \Bigr \} 
{\rm for} \ \ G_m=L_m, \ ML_m
\label{t21lm}
\eeqs

\beqs
N_{CSSG}(L_m) & = & N_{CSSG}(ML_m)-4m-1 \cr\cr
              & = & -2(m+1)+ \biggl [1+m \Bigl (1- \frac{1}{\sqrt{41}} \Bigr ) 
\biggr ]\Bigl (\frac{7+\sqrt{41}}{2} \Bigr )^m \cr\cr & & 
+ \biggl [1+m \Bigl (1+ \frac{1}{\sqrt{41}} \Bigr )
\biggr ]\Bigl (\frac{7-\sqrt{41}}{2} \Bigr )^m
\label{t12lm}
\eeqs

\beq
N_{SSG}(L_m)=N_{SSG}(ML_m) = 2^{e(L_m)}=2^{5m} \ .
\label{t22lm}
\eeq

Since $T(G_m,x,y)$ grows exponentially as $m
\to \infty$ for the families $G_m=S_m$ and $L_m$ for $(x,y)=(1,1)$,
(2,1), (1,2), and (2,2), it is natural to define the corresponding constants
\beq
z_{set}(\{G\}) = \lim_{n(G) \to \infty} n(G)^{-1} \ln N_{set}(G) \ , \quad
set = ST, \ SF, \ CSSG, \ SSG
\label{zset}
\eeq
where, as above, the symbol $\{G\}$ denotes the limit of the graph family $G$
as $n(G) \to \infty$ (and the $z$ here should not be confused with the
auxiliary expansion
variable in the generating function (\ref{gammatfbc}) or the Potts partition
function $Z(G,q,v)$.))  General inequalities for these were given in \cite{a}.
Our results yield
\beq
z_{ST}(\{G\}) = \frac{1}{4}\ln ( 3 + 2\sqrt{2} \ ) \simeq  0.440687
\quad {\rm for} \quad G = S, L, ML
\label{zst}
\eeq
\beq 
z_{SF}(\{G\}) = \frac{1}{4}\ln \Bigl [ 4 (4+\sqrt{15} \ ) \Bigr ] 
\simeq 0.862433 \quad {\rm for} \quad G=S,L,ML
\label{zsfzcssg}
\eeq
\beq
z_{CSSG}(\{G\}) = \frac{1}{4}\ln \Biggl ( \frac{ 7 + \sqrt{41}}{2} \Biggr ) 
\simeq 0.475585 \quad {\rm for} \quad G=S,L,ML
\label{zcssg}
\eeq
and
\beq
z_{SSG}(\{G\}) = (5/4)\ln 2 \simeq 0.866434 \quad {\rm for} \quad G=S,L,ML
\label{zssg}
\eeq
Another comparison of interest is the ratio of $z_{ST}$ for these $L_y=2$
strips with $z_{ST}$ for the full 2D honeycomb lattice, which has the value
\cite{wu77}
\beq
z_{ST}(hc)=\frac{1}{2}z_{ST}(tri) = 
\frac{3\sqrt{3}}{2\pi}(1 - 5^{-2} + 7^{-2} - 11^{-2} + 13^{-2} - ...) = 
0.807664868...
\label{zhcval}
\eeq
namely, 
\beq
\frac{z_{ST}(\{L\})}{z_{ST}(hc)} \simeq 0.545631 \ . 
\label{zstratio}
\eeq
In the case of
$z_{ST}$, it is also of interest to compare the results with a general upper
bound \cite{bbook}
\beq
z_{ST}(\{G\}) \le z_{ST,upper}(\{G\}) = \ln \Delta_{eff}(\{G\})
\label{zstupper}
\eeq
where the effective coordination number (degree) was defined above in eq. 
(\ref{delta_eff}).  For this comparison we define the ratio
\beq
r_{ST}(\{G\}) = \frac{z_{ST}(\{G\})}{z_{ST,upper}(\{G\})} \ . 
\label{rst}
\eeq
We find
\beq
r_{ST}(\{G\}) = \frac{\ln(3+2\sqrt{2})}{4\ln (\frac{5}{2}) } \ . 
\quad {\rm for} \quad G=S,L,ML
\label{rstg}
\eeq
We compare these results with others that we have obtained for infinite-length
strips of other lattices \cite{a,ta} in Table \ref{zvalues}.  We recall that
the strip of the triangular lattice is constructed by starting with a strip of
the square lattice and adding edges joining the lower left to upper right 
vertex of each square.  For uniformity,
when comparing the values of $z_{ST}(\{G\})$ to upper bounds, we use 
the general bound (\ref{zstupper}).  It should be remarked, however, that in
contrast to the strip graphs of the honeycomb lattice considered here, the
cyclic $L_y=2$ strip graphs of the square and triangular lattice and the
infinite-length open strips of these lattices are $\Delta$-regular, and 
hence one can apply a somewhat more restrictive upper bound, as discussed in 
\cite{st}. 

\begin{table}[hbtp]
\caption{\footnotesize{$z_s(\{G\})$ for the $L_x \to \infty$ limit of the width
$L_y=2$ strips of the (i) honeycomb lattice, (ii) square lattice, and (iii) 
triangular lattice.}}
\begin{center}
\begin{tabular}{|c|c|c|c|}
\hline
\hline
$z_s(\{G\})$      & $\{G\}=hc$  & $\{G\}=sq$  & $\{G\}=tri$ \\
\hline
$z_{ST}(\{G\})$   & $(1/4)\ln(3+2\sqrt{2} \ )$ 
                  & $(1/2)\ln(2+\sqrt{3} \ )$
                  & $(1/2)\ln[(7+3\sqrt{5} \ ) /2]$ \\ 
                  & $=0.440687$ & $=0.658479$ & $=0.962424$ \\

$r_{ST}(\{G\})$   & 0.4809 & 0.5994 & 0.6942 \\
\hline
$z_{SF}(\{G\})$   & $(1/4)\ln[4(4+\sqrt{15} \ )]$ 
                  & $(1/2)\ln[2(2+\sqrt{3} \ )]$
                  & $(1/2)\ln[2(3+2\sqrt{2} \ )]$ \\ 
                  & $=0.862433$ & $=1.005053$ & $=1.227947$ \\
\hline
$z_{CSSG}(\{G\})$ & $(1/4)\ln[(7+\sqrt{41} \ )/2]$ 
                  & $(1/2)\ln[(5+\sqrt{17}\ )/2]$
                  & $(1/2)\ln[2(3+2\sqrt{2} \ )]$ \\ 
                  & $=0.475585$ & $=0.758832$ & $=1.227947$ \\
\hline
$z_{SSG}(\{G\})$  & $(5/4)\ln 2$ 
                  & $(3/2)\ln 2$ 
                  & $2\ln 2$ \\ 
                  & $=0.866434$ & $=1.039721$ & $=1.386294$ \\
\hline
\end{tabular}
\end{center}
\label{zvalues}
\end{table}

The (planar) dual of $L_m$ can be described as follows.  In mathematical graph
theory, the ``join'' $G+H$ of the graphs $G$ and $H$ is defined as the graph
with the vertex set $V=V_G+V_H$ and edge set $E$ comprised of the edges of $G$
and $H$ together with edges joining each vertex of $G$ to each vertex of $H$.
Further, the complete graph $K_p$ is defined as the graph with $p$ vertices
such that each vertex is connected by an edge to every other vertex.  The
complement $\bar K_p$ of $K_p$ is the graph composed of $p$ disjoint vertices
and no edges.  Now consider the join $\bar K_2 + C_m$, where $C_m$ is the
circuit graph with $m$ vertices.  Change each of the edges connecting the two
vertices of the $\bar K_2$ to the vertices of the circuit graph $C_m$ to double
edges.  This yields a multigraph that we shall denote as $\bar K_2 (+)_{e2}
C_m$, where the symbol $G (+)_{e \ell} H$ means that one replaces each edge 
joining a vertex of $G$ with a vertex of $H$ by $\ell$ such edges. 
Then the (planar) dual of $L_m^*$ is 
\beq
L_m^* =  \bar K_2 (+)_{e2} C_m \ .
\label{lmdual}
\eeq
The planar dual of $S_m$
is the graph composed of the join of a single vertex $\bar K_1$ with the line
graph $T_m$ modified so that each edge connecting the external vertex to the 
$(m-2)$ internal vertices of the line graph is replaced by a 4-fold multiple 
edge, and the edges connecting this external vertex to the two end vertices 
of the line graph are replaced by 5-fold multiple edges. 
Thus, our calculations of various graph-theoretic quantities for $L_m$ and
$S_m$ also determine those for $L_m^*$ and $S_m^*$ by
eq. (\ref{tuttedual}). Note that this relation does not apply for the M\"obius
strips since they are not planar.

\vfill
\eject
\end{document}